\documentclass[preprint,prd,aps,showpacs,floatfix]{revtex4}
\begin{document}

%
 
\let\a=\alpha      \let\b=\beta       \let\c=\chi        \let\d=\delta   
\let\e=\varepsilon \let\f=\varphi     \let\g=\gamma      \let\h=\eta       
\let\k=\kappa      \let\l=\lambda     \let\m=\mu               
\let\o=\omega      \let\r=\varrho     \let\s=\sigma    
\let\t=\tau        \let\th=\vartheta  \let\y=\upsilon    \let\x=\xi       
\let\z=\zeta       \let\io=\iota      \let\vp=\varpi     \let\ro=\rho        
\let\ph=\phi       \let\ep=\epsilon   \let\te=\theta  
\let\n=\nu  
\let\D=\Delta   \let\F=\Phi    \let\G=\Gamma  \let\L=\Lambda  
\let\O=\Omega   \let\P=\Pi     \let\Ps=\Psi   \let\Si=\Sigma  
\let\Th=\Theta  \let\X=\Xi     \let\Y=\Upsilon  
 
%
 
%
 
\def\cA{{\cal A}}                \def\cB{{\cal B}} 
\def\cC{{\cal C}}                \def\cD{{\cal D}} 
\def\cE{{\cal E}}                \def\cF{{\cal F}} 
\def\cG{{\cal G}}                \def\cH{{\cal H}}                  
\def\cI{{\cal I}}                \def\cJ{{\cal J}}                 
\def\cK{{\cal K}}                \def\cL{{\cal L}}                  
\def\cM{{\cal M}}                \def\cN{{\cal N}}                 
\def\cO{{\cal O}}                \def\cP{{\cal P}}                 
\def\cQ{{\cal Q}}                \def\cR{{\cal R}}                 
\def\cS{{\cal S}}                \def\cT{{\cal T}}                 
\def\cU{{\cal U}}                \def\cV{{\cal V}}                 
\def\cW{{\cal W}}                \def\cX{{\cal X}}                  
\def\cY{{\cal Y}}                \def\cZ{{\cal Z}} 
%
 
\newcommand{\Ns}{N\hspace{-4.7mm}\not\hspace{2.7mm}} 
\newcommand{\qs}{q\hspace{-3.7mm}\not\hspace{3.4mm}} 
\newcommand{\ps}{p\hspace{-3.3mm}\not\hspace{1.2mm}} 
\newcommand{\ks}{k\hspace{-3.3mm}\not\hspace{1.2mm}} 
\newcommand{\des}{\partial\hspace{-4.mm}\not\hspace{2.5mm}} 
\newcommand{\desco}{D\hspace{-4mm}\not\hspace{2mm}}



\title{Democratic Mass Matrices From Five Dimensions} 
\author{A. Soddu}
\email{as7yf@galileo.phys.virginia.edu} 
\author{N-K. Tran} 
\email{nt6b@galileo.phys.virginia.edu}
\affiliation{Dept. of Physics, University of Virginia, \\
382 McCormick Road, P. O. Box 400714, Charlottesville, Virginia 22904-4714}
\date{\today}
\begin{abstract}

We reconstruct the Standard Model (SM) quark masses and the 
Cabibbo-Kobayashi-Maskawa (CKM) matrix from 
a five-dimensional model, with the fifth dimension compactified on an 
$S^1/Z_2$ orbifold. Fermions are localized only at the orbifold fixed 
points and the induced quark mass matrices are almost democratic. 
Two specific versions of our model with 15 and 24 parameters are presented, 
and for both versions we can 
reproduce the quark mass spectrum and CKM matrix correctly to the level 
they are observed in current experiments.
 
\end{abstract}

\maketitle
\section{Introduction}

The Standard Model (SM) has been the most satisfactory and widely recognized 
theory of particle interaction. To an extent, this is an effective theory, 
whose certain parameters are estimated and then refined by increasing 
high-precision experiments. However, this also means that the dynamical 
origin of some of these parameters is not found within the SM. One example 
is the pattern of family mixing characterized 
by the Cabibbo-Kobayashi-Maskawa (CKM) matrix, which in turn is related to 
the SM 
fermion mass spectrum and CP violation. Recently, considerable attention has 
been directed to phenomenological models with extra dimensions since among 
many other things, they can offer potential answers to puzzling 
questions related to the SM. 

In theories with compact extra dimensions, each original field in higher 
dimensional space can be effectively ``viewed'' as a tower of Kaluza-Klein (KK) 
states in equivalent 4-d theories after compactification processes. If SM 
is assumed to be the low-energy manifestation of a higher dimensional theory, 
the tower's lowest state (or KK zero mode) is identified as a SM field. In 
\cite{GGH} (see also earlier works \cite{JR,AS}), by introducing a 
$Z_2$-invariant Yukawa interaction 
between a background scalar field and a fermion field in 5-d theory, after a 
$ S^1/ Z_2$ compactification, one can obtain a non-trivial 
(i.e. localized) solution for the KK zero-mode wavefunction along the fifth 
dimension (brane scenario). From a 4-d point of view, any interaction term is 
now associated with 
a coupling being the overlap integration of extra dimensional 
wavefunctions of related fields. One then can flexibly control this 4-d 
theory effective coupling by regulating the localized wavefunctions along 
the extra dimension. This very interesting mechanism has found potential 
application in many problems such as proton decay suppression 
\cite{AS,MPST}, fermion 
mass hierarchy \cite{KT,GP,HS,HSS,MS}, CP violation \cite{BGR,CN}, etc.

In this work, we discuss the problem of quark mass spectrum and mixing 
angles by making extensive use of a democratic structure for the mass 
matrices. We found that a democratic structure for the mass matrices (DMM) is a 
particularly convenient choice, 
because it raises quite naturally in brane picture, and it can adequately 
generate both quark mass spectrum and CKM matrix to the precision 
determined by current experimental data. Previous works dealing with fermion 
mass hierarchy and CP violation within split-fermion scenario 
\cite{GP,MS,BGR,delAS,CN} focus on 
placing fermion families on different positions in the bulk, with the possibility to 
make mass matrix elements, originating from couplings between geographically very 
distant families, approximately zero. These highly hierarchical mass matrix 
approach however requires additional techniques (see section II) because naively 
localizing fermions in an arbitrary position along the extra dimension may 
contradict $S^1/Z_2$ orbifold compactification being used. In 
the DMM approach, we can avoid this subtlety by localizing all fields only at 
the two fixed points of the orbifold, while we break the family symmetry 
by modifying the detailed shapes of their wavefuctions. 
Further, the approach turns out to be more symmetric too.

Our work is structured as follows: in section II we introduce the Lagrangian 
to generate pure-phase mass matrices (PPMM) and DMM within a fermion 
localization mechanism, in section III we give the description of parameter 
space, in section IV we present the numerical method and report the results 
for the mass spectrum and CKM matrix for the 24 parameter version of our model and in section V 
for the 15 parameter version. Finally we give a brief conclusion 
of the work in section VI. 
 
\section{Formalism}
\subsection{Fermion Localization Mechanism}
We first briefly review the mechanism of fermion localization in extra 
dimension \cite{JR,AS,GGH}. We begin with the 5-d Lagrangian 
for a single massless fermion interacting with a real background scalar field. 
The fifth dimension is compact with support $[0,L]$. 
The generalization to the case with different families of fermions is 
straightforward.
         \begin{eqnarray}
{\cal L} =&& \bar{\psi} (x,y) (i\gamma ^{\mu} \partial _{\mu} - 
\gamma ^{5} \partial _{y} - f\phi (x,y)) \psi (x,y) \nonumber \\ 
&&+\frac{1}{2} \partial ^{\mu} \phi (x,y) \partial _{\mu} \phi (x,y) -  
\frac{1}{2} \partial _{y} \phi (x,y) \partial _{y} \phi (x,y) - 
\frac{\lambda}{4} (\phi ^2 (x,y) - V^2 )^2 \,.
         \label{Lag}
         \end{eqnarray} 
It is important that this Lagrangian is invariant under a $Z_2$ 
symmetry 
\footnote{A bare mass term of fermion is forbidden by this symmetry.}
         \begin{equation}
\phi (x,y) \rightarrow \Phi (x,y) \equiv - \phi (x,L-y) \,,
         \label{Z2Phi}
         \end{equation} 
         \begin{equation}
\psi (x,y) \rightarrow \Psi (x,y) \equiv \gamma _{5} \psi (x,L-y) \,,
         \label{Z2Psi}
         \end{equation}
To obtain the chiral zero mode for fermions as they are in the SM, it is 
necessary to compactify the 5-d theory on an $S^1/Z_2$ orbifold 
through the imposition of the following relations on the fields
         \begin{equation}
\phi (x,-y) = \Phi (x,L-y) =  \phi (x,2L-y) \,,
         \label{OBCPhi}
         \end{equation}
         \begin{equation}  
\psi (x,-y) = \Psi (x,L-y) =  \psi (x,2L-y) \,,
         \label{OBCPsi}
         \end{equation}
which constrain the fermion left-handed component and the scalar 
wavefunctions to be antisymmetric, and the fermion right-handed  component 
to be symmetric at the 
orbifold fixed points $y=0, L$. This in turn gives rise to a stable and 
non-constant VEV solution for the background scalar field when $L$ is 
sufficiently large ($L > \sqrt{\frac{1}{\lambda V^2}}$)
         \begin{equation}        
         \label{compositesol}
<\phi(x,y)>=h(y) = V \tanh {\left( \mu \frac{y}{L} \right )} 
\tanh {\left (  \mu(1-\frac{y}{L}) \right )} + 
{\cal{O}} (e^{-\mu}) \,,
         \end{equation} 
where $\mu \equiv L\sqrt{\frac{\lambda V^2}{2}} $ characterizes the extent 
of the brane in the transverse direction, to which Standard Model chiral quarks 
are going to be confined. Clearly, VEV kink-antikink approximation 
(\ref{compositesol}) holds only for $\mu > 1$. After performing 
a chiral decomposition 
         \begin{equation}
\psi (x,y) = \psi _R (x) \xi _R (y) +  \psi _L (x) \xi _L (y) \,,
         \label{chiraldecomp}
         \end{equation}
one obtains the massless (zero-mode) fermion wavefunctions satisfying the 
motion equation associated with Lagrangian (\ref{Lag})
         \begin{eqnarray}
         \label{solR}
\xi_{R}(y) &=&\frac{1}{N_R} \exp{\left(-f \int_{0}^{y} h(y')dy' \right)} = 
\frac{1}{N_R} \exp{\left[ F \left(\mu y + 
\frac{1}{\tanh \mu}\ln{\frac{\cosh \mu (1-y)}{\cosh \mu y \cosh \mu}}\right) 
\right]} \, ,
 \\
         \label{solL}
\xi_{L}(y) &=&\frac{1}{N_L} \exp{\left(f \int_{0}^{y} h(y')dy' \right)} = 
\frac{1}{N_L}\exp{\left[-F \left(\mu y + 
\frac{1}{\tanh \mu}\ln{\frac{\cosh \mu (1-y)}{\cosh \mu y \cosh \mu}}\right) 
\right]} \, ,
         \end{eqnarray}  
where $F \equiv f\sqrt{\frac{2}{\lambda}}$ and $N_{L,R}$ are the 
normalization factors. 

Now let us mention some important 
properties of these zero-mode solutions which are relevant to the present 
work. 

First, both solutions (\ref{solR}) and  (\ref{solL}) are 
symmetric at the orbifold fixed points $y=0,L$, so 
the right component survives and the left component vanishes 
by contradiction with the orbifold boundary conditions 
(\ref{OBCPsi}) assuring the single-handedness of SM fermions. 
By inverting the sign of $\gamma _5$ in (\ref{Z2Psi}) one can change the 
chirality of the surviving fermion. 

Second, the sign of the Yukawa 
coupling $f$ decides the localization position of the surviving chiral 
zero-mode fermions along the extra dimension; if $f>0$ it is $y=0$, 
if $f<0$ it is $y=L$. Localizing fermions at an arbitrary location 
other than the two fixed points in the bulk requires additional extensions of 
the Lagrangian such as odd-mass terms \cite{KT} or two background scalar fields 
\cite{GP}. In the next section, sticking just to the minimal localization 
mechanism, we exhaust all possibilities of placing different $SU(2)_W$ 
representations $Q$, $U$ and $D$ at the two fixed points and find that, in all 
configurations, this is indeed sufficient to obtain the right quark mass 
spectrum and CKM matrix. 
Remarkably, this minimal localization mechanism also features a democratic 
structure for the quark mass matrices, because fields of identical 
$SU(2)_W\times U(1)_Y$ gauge symmetry ($Q_{i}$'s ,$U_{i}$'s or $D_{i}$'s) 
are localized at the same point along the extra dimension. Small deviations 
from a democratic mass matrix, which is necessary in any realistic model, 
are realized in our approach by slightly modifying the fermions wavefunctions.  

Third, in the difference with other works in literature, here we 
make use of the exact solution for the zero-mode wavefunctions (\ref{solR}), 
(\ref{solL}) in place of a Gaussian profile approximation.
 
\subsection{Quark Flavor Mixing}
In the spirit of SM, we now introduce three $SU(2)_W$ doublets 
$Q_i$ and six $SU(2)_W$ singlets $U_i$, $D_i$ ($i=1,2,3$) whose zero 
modes are identified respectively with the SM quark chiral 
components $q_i$, $u_i$, $d_i$ after orbifold compactification. The Higgs doublet zero 
mode is assumed to be uniform along transverse direction 
($H(x,y)=H(x)/\sqrt{L}$). 

We construct a general 5-d 
Lagrangian concerning three fermion families 
        \begin{eqnarray}
{\cal L}_{f}&=&\sum_{i=1}^{3}  \bar{Q}_{i}(x,y)i\desco_5 Q_{i}(x,y) + 
(Q_{i} \leftrightarrow U_{i}) + (Q_{i} \leftrightarrow D_{i}) \\ \nonumber
&+&\sum_{i,j=1}^{3}\kappa^{u}_{5,ij}\bar{Q}_{i}(x,y) i\sigma_2 
H^{*}(x,y) U_{j}(x,y) + 
\sum_{i,j=1}^{3}\kappa^{d}_{5,ij}\bar{Q}_{i}(x,y)
H(x,y) D_{j}(x,y) + h.c. \, . 
        \end{eqnarray} 
In order to obtain a  PPMM in 4-d effective theory, 
we use the following ansatz for 5-d Yukawa couplings
        \begin{equation}
\kappa^{u}_{5,ij}=\kappa^{u}_{5}\exp{(i\theta^{u}_{ij})} \;\;\;\;\;\;\;\;\;\;
\kappa^{d}_{5,ij}=\kappa^{d}_{5}\exp{(i\theta^{d}_{ij})} \, ,
        \end{equation}
with $\kappa^{u}_{5,ij}$ and $\kappa^{d}_{5,ij}$ real positive.  
We note in particular in the above ansatz that complex higher-dimensional 
Yukawa couplings have universal absolute values 
$\kappa_{5}^{u}$, $\kappa_{5}^{d}$ for both up and down sector, and family 
symmetry is broken only in the phases. 
The difference between $\kappa^{u}_{5}$ and $\kappa^{d}_{5}$ to eventually 
give rise to the up-down quark mass hierarchy can also be accommodated 
conveniently in the brane picture with more extra dimensions (section IIC). 
The SM quarks obtain 
masses via spontaneous symmetry breaking with the Higgs developing a VEV 
$H(x,y) \rightarrow  (0,v/\sqrt{2L})^T $ (note that SM chiral fields 
$q_{ui}$ and $q_{di}$ have identical extra-dimensional wavefunctions 
because they originally come from the same doublet in higher-dimensional 
 theory) 
\footnote{As long as $v/\sqrt{2}\simeq 175 GeV < 1/L $, the Higgs zero mode is the only mode that 
receives non-zero VEV \cite{MPR} } 
        \begin{eqnarray}
\int dy {\cal L}_{f}&\rightarrow &
\sum_{i=1}^{3}  \bar{q}_{ui}(x)i\des q_{ui}(x) + 
(q_{ui} \leftrightarrow q_{di})+
(q_{ui} \leftrightarrow u_{i}) + (q_{ui} \leftrightarrow d_{i}) \\ \nonumber
&+&\sum_{i,j=1}^{3}[ \bar{q}_{ui}(x) M^u_{ij} u_{j}(x) + 
 \bar{q}_{di}(x) M^d_{ij} d_{j}(x)] + h.c. \, ,
        \end{eqnarray}
where 
        \begin{eqnarray}
         \label{massupnew}
M^u_{ij} = \frac{v}{\sqrt{2}}g_{Yu}\exp{(i\theta^{u}_{ij})} 
\int dy \xi_{qi}(y) \xi_{uj}(y) \, ,\\
M^d_{ij} =\frac{v}{\sqrt{2}} g_{Yd}\exp{(i\theta^{d}_{ij})} 
\int dy \xi_{qi}(y) \xi_{dj}(y) \, ,
         \label{massdownnew}
        \end{eqnarray}  

\noindent
with real, dimensionless effective couplings $g_{Yu}=\k_5^u/\sqrt{L}$ and 
$g_{Yd}=\k_5^d/\sqrt{L}$. 
One has to notice that a  pure-phase structure for the matrices $M^u$ and $M^d$ arises  
when the left-handed zero mode of $Q_i$  
are localized at the same position along the extra dimension independently of 
the family index $i$ and the same happens for the right-handed zero mode 
of $U_i$ and $D_i$. Then as it is clear from Eqs. (\ref{massupnew}) and (\ref{massdownnew}) 
the elements of each matrix differ only by a phase factor. 

We first 
perform the usual transformation from gauge eigenbasis to mass eigenbasis
        \begin{eqnarray}
u_{i}=U_{Rij}^{u} {u'}_{j}\, , &\;\;\;\;\;\;\;\;\;\;&
d_{i}=U_{Rij}^{d} {d'}_{j}\, ,
\\
q_{Li}^{u}=U_{Lij}^{u} {q'}_{Lj}^{u} \, , &\;\;\;\;\;\;\;\;\;\;&
q_{Li}^{d}=U_{Lij}^{d} {q'}_{Lj}^{d}\, ,
        \end{eqnarray}  
where $U_{L}^u$, $U_{L}^d$ diagonalize respectively the matrices ($M^u{M^u}^{\dagger}$), 
($M^d{M^d}^{\dagger}$) 
        \begin{eqnarray}
diag(|m_{u}|^2,|m_{c}|^2,|m_{t}|^2)=U_{L}^{u\dagger}
(M^{u}{M^u}^{\dagger})U_{L}^{u} \, , \\
diag(|m_{d}|^2,|m_{s}|^2,|m_{b}|^2)=U_{L}^{d\dagger}
(M^{d}{M^d}^{\dagger})U_{L}^{d} \, , 
        \end{eqnarray}  
and whose product gives the CKM matrix
        \begin{equation}
V_{CKM}=U_{L}^{u\dagger}U_{L}^{d} \, .
        \end{equation}
The origin of CP violation in weak interaction is related 
to the phase appearing in the CKM matrix, which by virtue of above relations 
comes from the complexity of mass matrices (it is well-known that 
real mass matrices do not give rise to CP violation). The 
localization mechanism in 5-d theory clearly provides a 
direct control over the modulus of each mass matrix element, but it does 
nothing to their phases. That is, a priori $M^u$ and  $M^d$ may 
possess in total 18 arbitrary phases $\theta^u_{ij}$, $\theta^d_{ij}$. 
\footnote{By rotating the right-handed quark fields one can absorb 3 
phases from each matrix, bringing to 12 the total number of phases.}
In the present work, stemming from the interest in model's simplicity, 
we just attribute four phases $\phi_{u1}$, $\phi_{u2}$, $\phi_{d1}$, 
$\phi_{d2}$ to elements of $M^u$ (say $M^u_{12}$, $M^u_{23}$) and  
$M^d$ (say $M^d_{12}$, $M^d_{23}$). 
One can note here that hermitian matrices ($M^u{M^u}^{\dagger}$) 
and ($M^d{M^d}^{\dagger}$) have altogether six phases, but two of them are 
non-physical and can be eliminated by a simultaneous transformation 
involving a single diagonal 
phase matrix $K=diag(1,\exp{(i\alpha)},\exp{(i\beta)})$ \cite{BGR}
        \begin{eqnarray}
M^u{M^u}^{\dagger} &\rightarrow& KM^u{M^u}^{\dagger}K^{\dagger} \, ,\\
M^d{M^d}^{\dagger} &\rightarrow& KM^d{M^d}^{\dagger}K^{\dagger} \, .
        \end{eqnarray}
Now the 4 physical phases left in $M^u{M^u}^{\dagger}$, $M^d{M^d}^{\dagger}$ can 
be reproduced by the chosen configuration with four phases in $M^u$, $M^d$.  

Let us next consider the magnitude of mass matrix elements, whose complete 
expressions are
        \begin{eqnarray}
        \label{elementMu}
M^u_{ij}& = & \frac{v}{\sqrt{2}}g_{Yu}\exp{(i\theta^u_{ij})} 
\int dy \xi_{qi}(y) \xi_{uj}(y)=
\nonumber \\ 
&& \frac{\exp{(i\theta^u_{ij})}}{N_{qi}N_{uj}}\int dy \exp\left[ F_{qi} 
\left(\mu_{qi} y + 
\frac{}{\tanh \mu_{qi}} \ln{\frac{\cosh \mu_{qi} (1-y)}
{\cosh \mu_{qi} y \cosh \mu_{qi}}}\right)\right. \nonumber \\ 
&&
\left. + F_{uj} \left(\mu_{uj} y + 
\frac{1}{\tanh \mu_{uj}} \ln{\frac{\cosh \mu_{uj} (1-y)}
{\cosh \mu_{uj} y \cosh \mu_{uj}}}\right) \right] \, , \\ 
M^d_{ij}& = & \frac{v}{\sqrt{2}}g_{Yd}\exp{(i\theta^d_{ij})} 
\int dy \xi_{qi}(y) \xi_{dj}(y)=
\nonumber  \\ 
&& \frac{\exp{(i\theta^d_{ij})}}{N_{qi}N_{dj}}
\int dy \exp\left[ F_{qi} \left(\mu_{qi} y + 
\frac{1}{\tanh \mu_{qi}} \ln{\frac{\cosh \mu_{qi} (1-y)}
{\cosh \mu_{qi} y \cosh \mu_{qi}}}\right)\right. \nonumber \\
&&
\left. + F_{dj} \left(\mu_{dj} y + 
\frac{1}{\tanh \mu_{dj}} \ln{\frac{\cosh \mu_{dj} (1-y)}
{\cosh \mu_{dj} y \cosh \mu_{dj}}}\right) \right] \, ,       
        \label{elementMd}
        \end{eqnarray}
where only $\theta^u_{12}$, $\theta^u_{23}$, $\theta^d_{12}$, 
$\theta^d_{23}$ are the non-zero phases. Indeed, without these phases, 
the solution satisfying the quark mass ratios and the 
CKM matrix cannot be obtained \cite{TS}. 
Further, if these phases are small enough, the 
mass matrices' structure transforms from almost pure-phase 
$M \approx g_Yv/\sqrt{2}\{e^{i\theta_{ij}}\}$ 
to almost democratic $M \approx g_Yv/\sqrt{2} \{1\} $. 
All our numerical solutions obtained below indeed clearly reflects this 
democratic structure. 
One very important advantage of a DMM is that it has 
three eigenvalues of ``loose hierarchy'': (0, 0, $3g_Yv/\sqrt{2}$), 
and by slightly modifying the mass matrix elements from ``1'' one can reproduce the 
right mass spectrum and right CKM matrix. 
More specifically, because small differences 
in $F$'s and $\mu$'s induce small modification in the corresponding  
wavefunction profiles $\xi(y)$'s, Eqs. (\ref{elementMu}), (\ref{elementMd}),  
to recover realistic quark masses  we will attribute different values 
of $F_i$ and $\mu_i$ to different flavors (the quartic coupling $\l$ is kept universal). 
Meanwhile 
we preserve essential democratic structure by localizing fields from
each of the three groups (doublets, up and down-type singlets) 
at the same point along the extra dimension regardless of family index. 
Choosing identical signs for $F_{qi}$ ($F_{ui}$, $F_{di}$) for different indices $i$ 
one can fulfill this requirement.    

Our approach hence is different from that of \cite{MS} where wavefunctions 
of different chiral flavors are very carefully and distinctly constructed 
in the bulk
so that their overlaps render the correct mass spectrum. However, associated 
CKM matrix found therein  does not generate sufficient CP violation as to 
the level it is observed in meson rare decays, 
(as long as the model has only one extra dimensions), even when one assigns to each 
mass matrix element an arbitrary phase \cite{BGR}. In the present work a DMM 
structure will be the key point to overcome this difficulty. 

\subsection{Six-dimensional Model}
With the model with just one extra dimension presented in the previous sections, one 
can fit quark mass spectrum and CKM matrix all by twisting around the 
pure-phase and democratic structures of mass matrices. We in particular have 
employed two different 5-d Yukawa couplings $\kappa_{5}^{u}$, $\kappa_{5}^{d}$
($\kappa_{5}^{u}/\kappa_{5}^{d} \sim 60$, sections IV-V) 
to generate up-down quark mass hierarchy, whose nature was not seen 
directly within the framework of 5-d theory. In this subsection, 
for the purpose of completeness, we briefly mention a possible solution to 
this issue, which consists of adding another spatial dimension to the 
theory. 

Beginning with six-dimensional model, we can repeat the orbifold 
compactification procedure for the two extra dimensions, one after the other, 
to secure the single-chirality of zero modes. 
We choose to localize all doublets identically along the sixth dimension
(the same holds for up-type and down-type singlets). In the result, the 
5-d (now effective) Yukawa coupling $\kappa_{u}^{5}$ ($\kappa_{d}^{5}$) 
are just the product of 6-d couplings $\kappa_{u}^{6}$ ($\kappa_{d}^{6}$) 
and the $Q-U$ ($Q-D$) wavefunction overlap along the sixth dimension. So 
by starting with a single Yukawa coupling ($\kappa_{u}^{6}=\kappa_{d}^{6}$) 
in 6-d theory, we can end up with two different 5-d couplings because $U$ 
and $D$ fields have been placed differently from $Q$ fields along the sixth 
dimension. Further, the phases of mass matrices' elements could also be 
generated from 6-d models (see details in \cite{HS,HSS}). Rather, the point 
we would like to emphasize here is that, extra dimension theory can potentially 
provide necessary ingredients to reproduce 4-d effective theory of particle 
interaction.

\section{Description of the Parameter Space}

In this section we present the parameter space for the particular choice of the model 
considered with 24 parameters.  
Also if the number of parameters is large, what has to be said here is that the 
``naturality'' of the 
parameters, (generally all the values are of order one or differ by no more than one order of magnitude), 
we believe, is the most important factor, and the parameter space found satisfies this condition. 
The list of the parameters is the following:
    
\begin{itemize}
\item{$g_{Yu}$ and  $g_{Yd}$} 
\item{$\mu_{qi}=L\sqrt{\frac{{\lambda}V_{qi}^2}{2}}$, 
$\mu_{ui}=L\sqrt{\frac{{\lambda}V_{ui}^2}{2}}$,  and 
$\mu_{di}=L\sqrt{\frac{{\lambda}V_{di}^2}{2}}$, with $i=1,2,3$ are dimensionless 
quantities whose inverse 
is proportional to the thickness of the domain wall.}
\item{$F_{qi}=\sqrt{2/\l}f_{qi}$, $F_{ui}=\sqrt{2/\l}f_{ui}$ 
and $F_{di}=\sqrt{2/\l}f_{di}$ with $i=1,2,3$ and $f$'s being the Yukawa 
couplings appearing in Eq. (\ref{Lag})}
\item{$\phi_{u1}$, $\phi_{u2}$, $\phi_{d1}$, and $\phi_{d2}$ 
are the phases appearing respectively in the up and 
down mass matrices}
\end{itemize} 

As it can be seen from this particular choice of the parameter space, 
we decided to break family symmetry by choosing different values 
for $\mu_i$ and $f_i$ for different indices $i$ (toghether with four 
different phases $\theta^{u}_{12}$, $\theta^{u}_{23}$, $\theta^{d}_{12}$, 
$\theta^{d}_{23}$ appearing in the mass matrices), and at the same time 
to break the left-right symmetry by different values for the left 
component parameters $\mu_q$ and $f_q$ 
and the right component parameters $\mu_u$, $f_u$, $\mu_d$ and $f_d$. 

\section{Results for Mass Matrices from Five Dimensions}

In this section we present the numerical results obtained for the 
parameter space and for the physical quantities of Table 1. 
We consider four different cases, which correspond to the all 
four possible ways of picking  
the sign of the Yukawa couplings $f$ for the left and right components. 
The four different cases are the following:

\begin{itemize}
\item{$f_{qi}>0 \,\, f_{ui}>0 \,\, f_{di}>0$ denoted as $(+++)$}
\item{$f_{qi}>0 \,\, f_{ui}>0 \,\, f_{di}<0$ denoted as $(++-)$}
\item{$f_{qi}>0 \,\, f_{ui}<0 \,\, f_{di}>0$ denoted as $(+-+)$}
\item{$f_{qi}>0 \,\, f_{ui}<0 \,\, f_{di}<0$ denoted as $(+--)$}
\end{itemize}

\noindent
The first case corresponds to the doublets, up and down-type singlets all 
localized at the orbifold fixed point $y=0$. 
The second case corresponds to the doublets and up-type singlets localized 
at $y=0$, while the down-type singlets at $y=L$. 
The third case corresponds to the doublets and down-type singlets localized 
at $y=0$, while the up-type singlets at $y=L$. 
And finally the fourth case corresponds to the doublets localized 
at $y=0$, while both the up and down-type singlets at $y=L$. 

The other four possible cases obtained when one changes at the same time 
all the signs of the Yukawa couplings are just 
symmetrical to the four presented, with each wave function now localized at the 
other orbifold fixed point, and with symmetrical profile. So they do not 
present any new mixing pattern.    

The approach we use to derive the parameter space consists in minimizing a 
particular function, built in such a way that its global minima correspond 
to the region defined by the experimental constraints. 
This function is defined as:

\begin{eqnarray}
E & = & \sum_{i=1}^N \frac{(x^{th}_i-x^{min}_i)^2}{<x_i>^2}
\,\theta(x^{min}_i-x^{th}_i) \nonumber \\
& + & \sum_{i=1}^N \frac{(x^{th}_i-x^{max}_i)^2}{<x_i>^2}
\,\theta(x^{th}_i-x^{max}_i)
\label{EE}
\end{eqnarray} 

\noindent
where $\theta(x)$ is the step function, N is the number of quantities 
that we want to fit, $x^{th}_i$ is the 
predicted value for the $ith$ quantity, $x^{min}_i $ and $x^{max}_i$ 
fix the range for the $ith$ quantity, and $<x_i>$ is its average value.   
It is immediate to verify from Eq. (\ref{EE}) that when all the predicted 
quantities $x^{th}_i$'s are contained in the proper ranges, the function $E$ 
takes its minimum value equal to zero. 
The set of parameters which correspond to a zero value for the 
function $E$ is called a solution.

The minimization procedure we used is called 
simulated annealing \cite{SA1}\cite{SA2}, and when the function that we want to minimize 
depends on many parameters, this procedure seems to work more efficiently than others. 
In particular the simulated annealing method 
is mostly used when the global minima are surrounded by a lot of local minima. 
In fact this minimization process can find a global minimum also after being 
trapped in a local minimum.   

In the following we will present the numerical results for each of the cases mentioned above. 
We will present graphically together with the parameter space (Figs. 
\ref{Figure_1},\ref{Figure_6},\ref{Figure_11},\ref{Figure_16}), the solutions of the 
mass spectrum (Figs. \ref{Figure_3},\ref{Figure_8},\ref{Figure_13},\ref{Figure_18}) 
(masses are given in GeV and are evaluated at the $M_Z$ scale), 
the CKM matrix (Figs. 4,9,14,19), and the $\bar{\rho},\bar{\eta} $ CP parameters 
(Figs. \ref{Figure_5},\ref{Figure_10},\ref{Figure_15},\ref{Figure_20}).
For each case we will also give one particular numerical example of all model's parameters, the 
corresponding mass matrices with eigenvalues (i.e. quark masses), 
the rotation matrices, the CKM matrix, the CP parameters, the corresponding plots of background scalar 
fields and the wavefunction profiles for left and right components 
(Figs. \ref{Figure_2},\ref{Figure_7},\ref{Figure_12},\ref{Figure_17}). Complex phases are 
measured in radiant.  

\begin{itemize}
\item{$f_{qi}>0 \,\, f_{ui}>0 \,\, f_{di}>0$}
\end{itemize}

\begin{equation}
g_{Yu(24)}^{(+++)}\frac{v}{\sqrt{2}} =57.81 \, , \,\,\, 
g_{Yd(24)}^{(+++)}\frac{v}{\sqrt{2}} = 0.98 \, ,
\end{equation}
\begin{equation}
F_{q1(24)}^{(+++)} = 1.389 \, , \,\,\, F_{q2(24)}^{(+++)} = 0.979 \, , 
\,\,\, F_{q3(24)}^{(+++)}= 0.787 \, ,
\end{equation}
\begin{equation}
F_{u1(24)}^{(+++)} =0.938 \, , \,\,\, F_{u2(24)}^{(+++)} =0.843 \, , 
\,\,\, F_{u3(24)}^{(+++)}= 1.352 \, ,
\end{equation}
\begin{equation}
F_{d1(24)}^{(+++)} = 1.344 \, , \,\,\, F_{d2(24)}^{(+++)} = 1.013 \, , 
\,\,\, F_{d3(24)}^{(+++)}= 1.437 \, ,
\end{equation}
\begin{equation}
\m_{q1(24)}^{(+++)} = 2.252 \, , \,\,\, \m_{q2(24)}^{(+++)} =3.367 \, , 
\,\,\, \m_{q3(24)}^{(+++)}= 2.660 \, ,
\end{equation}
\begin{equation}
\m_{u1(24)}^{(+++)} = 1.965 \, , \,\,\, \m_{u2(24)}^{(+++)} = 2.060 \, , 
\,\,\, \m_{u3(24)}^{(+++)}= 1.496 \, ,
\end{equation}
\begin{equation}
\m_{d1(24)}^{(+++)} = 2.537 \, , \,\,\, \m_{d2(24)}^{(+++)} = 3.157 \, , 
\,\,\, \m_{d3(24)}^{(+++)}= 2.520 \, .
\end{equation}
\begin{equation}
\phi_{u1(24)}^{(+++)} = -0.0001\, , \,\,\, \phi_{u2(24)}^{(+++)} =  0.0155\, ,  
\,\,\,\phi_{d1(24)}^{(+++)} =-0.0095 \, , \,\,\, \phi_{d2(24)}^{(+++)} = -0.1607\, ,
\end{equation}

\begin{equation}
M_{u(24)}^{(+++)}=57.81 \,GeV\,\left(\begin{array}{ccc}
0.9814 & \,0.9811 \,e^{-i\,0.0001}  & 0.9705 \\           
0.9443 & 0.9438 & \,0.9268\, e^{i\,0.0155}\\
0.9938 & 0.9936 & 0.9870  
\end{array}\right) \, ,
\label{matrixupPPP}
\end{equation}  
\begin{equation}
m_{u(24)}^{(+++)} = 0.0027 \,GeV\,,\,\,\,\, 
m_{c(24)}^{(+++)}=0.677 \,GeV\,,\,\,\,\, m_{t(24)}^{(+++)}=168.13 \,GeV\, ,
\label{eigenupPPP}
\end{equation}
\begin{equation}
M_{d(24)}^{(+++)}=0.975 \, GeV \,\left(\begin{array}{ccc}
0.9973 & \,0.9934\,e^{-i\,0.0095}   & 0.9955 \\           
0.9975 & 0.9996  & \,0.9988\,e^{-i\,0.1607}  \\
0.9880 & 0.9808 & 0.9845 
\end{array}\right) \, ,
\label{matrixdownPPP}
\end{equation}  
\begin{equation}
m_{d(24)}^{(+++)} = 0.0048 \,GeV\,,\,\,\,\, 
m_{s(24)}^{(+++)}=0.106 \,GeV\,,\,\,\,\, m_{b(24)}^{(+++)}=2.90 \,GeV\, .
\label{eigendownPPP}
\end{equation}

In Eqs. (\ref{matrixupPPP}), (\ref{matrixdownPPP}) the mass matrices are written in a form that better 
evidenciates the almost democratic structure.  
In Eqs. (\ref{UudagaPPP}) and (\ref{UdPPP}) we give the expressions for the rotation matrices 
$U_L^{u\dag}$ and $U_L^d$ whose product is just the CKM matrix, Eq. (\ref{CKMmatPPP}). 
In Eq. (\ref{absCKMmatPPP}) we give the absolute value of the CKM matrix, and 
in Eq. (\ref{rhoetavalPPP}) the values for the CP parameters $\bar{\rho}$ and $\bar{\eta}$ and 
the invariant area of the unitary triangle $J_{CP}$.  

\begin{equation}
U_{L(24)}^{u\dag (+++)}=\left(\begin{array}{ccc}
0.7429 &-0.1027 - 0.1315i  & -0.6361 + 0.1249i \\           
-0.3067 - 0.1223i & 0.8123 & -0.4664 + 0.1167i \\
0.5823 & 0.5588 - 0.0028i &  0.5905
\end{array}\right) \, ,
\label{UudagaPPP}
\end{equation}  

\begin{equation}
U_{L(24)}^{d(+++)}=\left(\begin{array}{ccc}
-0.6924 + 0.0002i  & -0.4298 - 0.0234i  & 0.5783 + 0.0292i  \\           
-0.0207 + 0.0031i & 0.8141 & 0.5803 \\
0.7212 & -0.3893 - 0.0191i  & 0.5718 + 0.0306i  
\end{array}\right) \, ,
\label{UdPPP}
\end{equation}

\begin{equation}
V_{CKM(24)}^{(+++)}=\left(\begin{array}{ccc}
-0.9706 + 0.0927i &-0.1529- 0.1609i  & 0.0025 - 0.0027i \\           
-0.1408+ 0.1713i & 0.9741 + 0.0232i & 0.0274 - 0.0272i \\
0.0111 + 0.0019i &-0.0252 - 0.0272i & 0.9987 + 0.0334i
\end{array}\right) \, ,
\label{CKMmatPPP}
\end{equation}  
   
\begin{equation}
\left|V_{CKM(24)}^{(+++)}\right|=\left(\begin{array}{ccc}
0.9750 &0.2220  & 0.0037 \\           
0.2217 & 0.9744 & 0.0386 \\
0.0113 &0.0371  & 0.9992 
\end{array}\right) \, ,
\label{absCKMmatPPP}
\end{equation}  

\begin{equation}
\bar{\rho}^{(+++)}_{(24)} = 0.28\, , \,\,\,\,\, \bar{\eta}^{(+++)}_{(24)} = 0.31\, , 
\,\,\,\,\, J_{CP(24)}^{(+++)}=-2.2\times 10^{-5} \, .
\label{rhoetavalPPP}
\end{equation}

\noindent
with $\bar{\rho}$ and $\bar{\eta}$ defined as

\begin{equation}
\bar{\rho} = Re(V_{ud}V_{ub}^*V_{cd}^*V_{cb})/|V_{cd}V_{cb}^*|^2 \, , 
\end{equation}
\begin{equation}
\bar{\eta} = Im(V_{ud}V_{ub}^*V_{cd}^*V_{cb})/|V_{cd}V_{cb}^*|^2 \, ,
\end{equation}

and $J_{CP}$ as

\begin{equation}
J_{CP} = Im(V_{us}V_{ub}^{*}V_{cs}^{*}V_{cb}) \, .
\end{equation}

%
%
\begin{itemize}
\item{$f_{qi}>0 \,\, f_{ui}>0 \,\, f_{di}<0$}
\end{itemize}

\begin{equation}
g_{Yu(24)}^{(++-)}\frac{v}{\sqrt{2}} = 59.85 \, , 
\,\,\, g_{Yd(24)}^{(++-)}\frac{v}{\sqrt{2}} = 1.18 \, ,
\end{equation}
\begin{equation}
F_{q1(24)}^{(++-)} = 0.482 \, , \,\,\, F_{q2(24)}^{(++-)} =0.888 \, , 
\,\,\, F_{q3(24)^{(++-)}}= 2.522 \, ,
\end{equation}
\begin{equation}
F_{u1(24)}^{(++-)} = 0.966 \, , \,\,\, F_{u2(24)}^{(++-)} = 0.725 \, , 
\,\,\, F_{u3(24)}^{(++-)}= 1.185 \, ,
\end{equation}
\begin{equation}
F_{d1(24)}^{(++-)} = -1.024 \, , \,\,\, F_{d2(24)}^{(++-)} = -1.639 \, , 
\,\,\, F_{d3(24)}^{(++-)}= -0.014 \, ,
\end{equation}
\begin{equation}
\m_{q1(24)}^{(++-)} = 2.279 \, , \,\,\, \m_{q2(24)}^{(++-)} = 2.474 \, , 
\,\,\, \m_{q3(24)}^{(++-)}= 1.234 \, ,
\end{equation}
\begin{equation}
\m_{u1(24)}^{(++-)} = 1.755 \, , \,\,\, \m_{u2(24)}^{(++-)} =2.597  \, , 
\,\,\, \m_{u3(24)}^{(++-)}= 2.000 \, ,
\end{equation}
\begin{equation}
\m_{d1(24)}^{(++-)} = 2.223 \, , \,\,\, \m_{d2(24)}^{(++-)} = 2.079 \, , 
\,\,\, \m_{d3(24)}^{(++-)}= 1.482 \, ,
\end{equation}
\begin{equation}
\phi_{u1(24)}^{(++-)} = -0.0007 \, , \,\,\, \phi_{u2(24)}^{(++-)} = -0.0018 \, ,  
\,\,\,\phi_{d1(24)}^{(++-)} = 0.0122 \, , \,\,\, \phi_{d2(24)}^{(++-)} =-0.0951 \, .
\end{equation}

\begin{equation}
M_{u(24)}^{(++-)}= 59.85\,GeV\,\left(\begin{array}{ccc}
0.9997 & \,0.9899\,e^{-i\,0.0007}  & 0.9910 \\                     
0.9887 & 0.9995 & \,0.9992\,e^{-i\,0.0018}  \\
0.9998 & 0.9952 & 0.9960
\end{array}\right) \, ,
\label{matrixupPPN}
\end{equation}  
\begin{equation}
m_{u(24)}^{(++-)} =  0.0024\,GeV\,,\,\,\,\, m_{c(24)}^{(++-)}=0.722 \,GeV\,,
\,\,\,\, m_{t(24)}^{(++-)}=178.7 \,GeV\, ,
\label{eigenupPPN}
\end{equation}
\begin{equation}
M_{d(24)}^{(++-)}= 1.18\, GeV \,\left(\begin{array}{ccc}
0.8867 & \,0.8337\,e^{i\,0.0122}   & 0.9863 \\           
0.7955 & 0.7293  & \,0.9428\,e^{-i\,0.0951}  \\
0.8656 & 0.8090 & 0.9779
\end{array}\right) \, ,
\label{matrixdownPPN}
\end{equation}  
\begin{equation}
m_{d(24)}^{(++-)} =  0.0051\,GeV\,,\,\,\,\, m_{s(24)}^{(++-)}= 0.082\,GeV\,,
\,\,\,\, m_{b(24)}^{(++-)}= 3.1\,GeV\, .
\label{eigendownPPN}
\end{equation}

\begin{equation}
U_{L(24)}^{u\dag (++-)}=\left(\begin{array}{ccc}
-0.5859 + 0.0191i  & -0.2010 - 0.0193i  & 0.7846 \\           
-0.5691 + 0.0193i & 0.7910 & -0.2229 - 0.0189i \\
0.5763 + 0.0001i & 0.5775 + 0.0003i &  0.5783
\end{array}\right) \, ,
\label{UudagaPPN}
\end{equation}  

\begin{equation}
U_{L(24)}^{d(++-)(24)}=\left(\begin{array}{ccc}
-0.6624 - 0.1144i & -0.4187 - 0.1241i  & 0.5979\\           
 -0.0511 + 0.1249i & 0.8265  & 0.5461 - 0.0235i \\
0.7280 & -0.3424 + 0.0946i  &  0.5863 - 0.0019i
\end{array}\right) \, ,
\label{UdPPN}
\end{equation}

\begin{equation}
V_{CKM(24)}^{(++-)}=\left(\begin{array}{ccc}
0.9741 + 0.0303i &-0.1870 + 0.1230i  & -0.0004 + 0.0041i   \\           
0.1765 + 0.1373i &0.9726 + 0.0479i  & -0.0390 - 0.0177i \\
0.0097 + 0.0061i & 0.0380 - 0.0166i & 0.9990 - 0.0144i
\end{array}\right) \, ,
\label{CKMmatPPN}
\end{equation}  
   
\begin{equation}
\left|V_{CKM(24)}^{(++-)}\right|=\left(\begin{array}{ccc}
0.9746 & 0.2239  & 0.0041 \\           
0.2236 & 0.9737 & 0.0428 \\
0.0115 & 0.0415 & 0.9991 
\end{array}\right) \, ,
\label{absCKMmatPPN}
\end{equation}  

\begin{equation}
\bar{\rho}^{(++-)}_{(24)} = 0.13\, , \,\,\,\,\, 
\bar{\eta}^{(++-)}_{(24)} =0.40 \, , \,\,\,\,\, J_{CP(24)}^{(++-)}=-3.6\times 10^{-5} \, .
\label{rhoetavalPPN}
\end{equation}

%
%

\begin{itemize}
\item{$f_{qi}>0 \,\, f_{ui}<0 \,\, f_{di}>0$}
\end{itemize}

\begin{equation}
g_{Yu(24)}^{(+-+)}\frac{v}{\sqrt{2}} = 61.66 \, , \,\,\, 
g_{Yd(24)}^{(+-+)}\frac{v}{\sqrt{2}} = 1.00 \, ,
\end{equation}
\begin{equation}
F_{q1(24)}^{(+-+)} = 3.152 \, , \,\,\, F_{q2(24)}^{(+-+)} =0.327 \, , 
\,\,\, F_{q3(24)}^{(+-+)}= 0.485 \, ,
\end{equation}
\begin{equation}
F_{u1(24)}^{(+-+)} =-0.112 \, , \,\,\, F_{u2(24)}^{(+-+)} = -0.629 \, , 
\,\,\, F_{u3(24)}^{(+-+)}= -0.516 \, ,
\end{equation}
\begin{equation}
F_{d1(24)}^{(+-+)} = 1.062 \, , \,\,\, F_{d2(24)}^{(+-+)} =0.087  \, , 
\,\,\, F_{d3(24)}^{(+-+)}= 1.244 \, ,
\end{equation}
\begin{equation}
\m_{q1(24)}^{(+-+)} = 1.467 \, , \,\,\, \m_{q2(24)}^{(+-+)} =2.142 \, , 
\,\,\, \m_{q3(24)}^{(+-+)}= 2.458 \, ,
\end{equation}
\begin{equation}
\m_{u1(24)}^{(+-+)} = 1.942 \, , \,\,\, \m_{u2(24)}^{(+-+)} = 1.658 \, , 
\,\,\, \m_{u3(24)}^{(+-+)}= 1.452 \, ,
\end{equation}
\begin{equation}
\m_{d1(24)}^{(+-+)} = 2.008 \, , \,\,\, \m_{d2(24)}^{(+-+)} = 1.820 \, , 
\,\,\, \m_{d3(24)}^{(+-+)}= 2.460 \, ,
\end{equation}
\begin{equation}
\phi_{u1(24)}^{(+-+)} = -0.0006 \, , \,\,\, \phi_{u2(24)}^{(+-+)} = -0.0000  \, ,  
\,\,\,\phi_{d1(24)}^{(+-+)} = 0.0870 \, , \,\,\, \phi_{d2(24)}^{(+-+)} = -0.0035 \, ,
\end{equation}

\begin{equation}
M_{u(24)}^{(+-+)}= 61.66\,GeV\,\left(\begin{array}{ccc}
0.9185 & \,0.8840\,e^{-i\,0.0006}  & 0.9032 \\           
0.9920 & 0.9785  & \,0.9865\,e^{-i\,0.0000} \\
0.9763 & 0.9557 & 0.9676
\end{array}\right) \, ,
\label{matrixupPNP}
\end{equation}  
\begin{equation}
m_{u(24)}^{(+-+)} = 0.0024 \,GeV\,,\,\,\,\, m_{c(24)}^{(+-+)}=0.724 
\,GeV\,,\,\,\,\, m_{t(24)}^{(+-+)}=176.1 \,GeV\, ,
\label{eigenupPNP}
\end{equation}
\begin{equation}
M_{d(24)}^{(+-+)}= 1.00\, GeV \,\left(\begin{array}{ccc}
0.9941 & \,0.9357\,e^{i\,0.0870}   & 0.9977 \\           
0.9851 & 0.9968  & \,0.9403\,e^{-i\,0.0035}  \\
0.9967 & 0.9853  & 0.9672
\end{array}\right) \, ,
\label{matrixdownPNP}
\end{equation}  
\begin{equation}
m_{d(24)}^{(+-+)} = 0.0048 \,GeV\,,\,\,\,\, m_{s(24)}^{(+-+)}= 0.082\,GeV\,,
\,\,\,\, m_{b(24)}^{(+-+)}= 2.9\,GeV\, .
\label{eigendownPNP}
\end{equation}

\begin{equation}
U_{L(24)}^{u\dag (+-+)}=\left(\begin{array}{ccc}
-0.2669 + 0.0067i &-0.5391 - 0.0062i  & 0.7988 \\           
0.7934 & -0.5933 - 0.0066i  & -0.1354 + 0.0069i \\
0.5470 + 0.0001i & 0.5977 & 0.5861
\end{array}\right) \, ,
\label{UudagaPNP}
\end{equation}  

\begin{equation}
U_{L(24)}^{d(+-+)}=\left(\begin{array}{ccc}
-0.1495 + 0.1160i & 0.7951   & 0.5759\\           
-0.6160 - 0.1196i & -0.5151 - 0.1001i  & 0.5753 - 0.0008i\\
0.7553 & -0.2785 + 0.1220i  &  0.5806 
\end{array}\right) \, ,
\label{UdPNP}
\end{equation}

\begin{equation}
V_{CKM(24)}^{(+-+)}=\left(\begin{array}{ccc}
0.9738 + 0.0363i & -0.1576 + 0.1600i & -0.0002 - 0.0036i \\           
0.1437 + 0.1723i & 0.9727 + 0.0444i & 0.0370 + 0.0135i \\
-0.0073 - 0.0081i & -0.0362 + 0.0118i & 0.9992 + 0.0084i
\end{array}\right) \, ,
\label{CKMmatPNP}
\end{equation}  
   
\begin{equation}
\left|V_{CKM(24)}^{(+-+)}\right|=\left(\begin{array}{ccc}
0.9744 & 0.2246 & 0.0036 \\           
0.2244 &0.9737  & 0.0394\\
0.0108 &0.0381  & 0.9992 
\end{array}\right) \, ,
\label{absCKMmatPNP}
\end{equation}  

\begin{equation}
\bar{\rho}_{(24)}^{(+-+)} = 0.17\, , \,\,\,\,\, \bar{\eta}_{(24)}^{(+-+)} = 0.36\, , 
\,\,\,\,\, J_{CP(24)}^{(+-+)}=-2.8\times 10^{-5} \, .
\label{rhoetavalPNP}
\end{equation}

%
%

\begin{itemize}
\item{$f_{qi}>0 \,\, f_{ui}<0 \,\, f_{di}<0$}
\end{itemize}

\begin{equation}
g_{Yu(24)}^{(+--)}\frac{v}{\sqrt{2}} = 60.83 \, , 
\,\,\, g_{Yd(24)}^{(+--)}\frac{v}{\sqrt{2}} = 1.25 \, ,
\end{equation}
\begin{equation}
F_{q1(24)}^{(+--)} = 2.233  \, , \,\,\, F_{q2(24)}^{(+--)} =0.357 \, , 
\,\,\, F_{q3(24)}^{(+--)}= 1.069 \, ,
\end{equation}
\begin{equation}
F_{u1(24)}^{(+--)} =-0.429 \, , \,\,\, F_{u2(24)}^{(+--)} = -1.476 \, , 
\,\,\, F_{u3(24)}^{(+--)}= -0.313 \, ,
\end{equation}
\begin{equation}
F_{d1(24)}^{(+--)} = -0.323 \, , \,\,\, F_{d2(24)}^{(+--)} = -0.847 \, , 
\,\,\, F_{d3(24)}^{(+--)}= -2.480 \, ,
\end{equation}
\begin{equation}
\m_{q1(24)}^{(+--)} = 2.087 \, , \,\,\, \m_{q2(24)}^{(+--)} =2.236 \, , 
\,\,\, \m_{q3(24)}^{(+--)}= 2.130 \, ,
\end{equation}
\begin{equation}
\m_{u1(24)}^{(+--)} = 1.126 \, , \,\,\, \m_{u2(24)}^{(+--)} = 1.006 \, , 
\,\,\, \m_{u3(24)}^{(+--)}= 1.322 \, ,
\end{equation}
\begin{equation}
\m_{d1(24)}^{(+--)} = 3.380 \, , \,\,\, \m_{d2(24)}^{(+--)} = 1.736 \, , 
\,\,\, \m_{d3(24)}^{(+--)}= 1.693 \, .
\end{equation}
\begin{equation}
\phi_{u1(24)}^{(+--)} = -0.0053 \, , \,\,\, \phi_{u2(24)}^{(+--)} = -0.0003 \, ,  \,\,\,
\phi_{d1(24)}^{(+--)} =0.0218 \, , \,\,\, \phi_{d2(24)}^{(+--)} =-0.0525 \, ,
\end{equation}

\begin{equation}
M_{u(24)}^{(+--)}= 60.83\,GeV\,\left(\begin{array}{ccc}
0.8427 & \,0.8188\,e^{-i\,0.0053}   & 0.8412\\           
0.9893 & 0.9819  & \,0.9889\,e^{-i\,0.0003}  \\
0.9447 & 0.9293  & 0.9437  
\end{array}\right) \, ,
\label{matrixupPNN}
\end{equation} 
\begin{equation}
m_{u(24)}^{(+--)} = 0.0022 \,GeV\,,\,\,\,\, m_{c(24)}^{(+--)}= 0.677\,GeV\,,
\,\,\,\, m_{t(24)}^{(+--)}= 168.3\,GeV\, ,
\label{eigenupPNN}
\end{equation}
\begin{equation}
M_{d(24)}^{(+--)}= 1.25\, GeV \,\left(\begin{array}{ccc}
0.7298 & \,0.7668\,e^{i\,0.0218} & 0.5944 \\           
0.9438 & 0.9613  & \,0.8640\,e^{-i\,0.0525} \\
0.8664 &0.8936  & 0.7580 
\end{array}\right) \, ,
\label{matrixdownPNN}
\end{equation}   
\begin{equation}
m_{d(24)}^{(+--)} = 0.0050 \,GeV\,,\,\,\,\, m_{s(24)}^{(+--)}= 0.081\,GeV\,,
\,\,\,\, m_{b(24)}^{(+--)}= 3.1\,GeV\, .
\label{eigendownPNN}
\end{equation}

\begin{equation}
U_{L(24)}^{u\dag (+--)}=\left(\begin{array}{ccc}
-0.3525 + 0.0856i  & -0.4668 - 0.0730i & 0.8033i \\           
0.7718 & -0.6230 - 0.0844i  & -0.0311 + 0.0898i  \\
0.5223 + 0.0008i & 0.6177  &  0.5880 - 0.0001i
\end{array}\right) \, ,
\label{UudagaPNN}
\end{equation}  

\begin{equation}
U_{L(24)}^{d(+--)}=\left(\begin{array}{ccc}
-0.4256 - 0.2346i & 0.7244  &0.4889 + 0.0110i \\           
-0.3737 + 0.1807i & -0.5963 + 0.2371i  & 0.6449\\
0.7691 & 0.0554 - 0.2459i  & 0.5873 + 0.0085i  
\end{array}\right) \, ,
\label{UdPNN}
\end{equation}

\begin{equation}
V_{CKM(24)}^{(+--)}=\left(\begin{array}{ccc}
0.9755 - 0.0108i & 0.0848 - 0.2026i & -0.0025 - 0.0022i  \\           
-0.1043 - 0.1930i & 0.9710 - 0.0847i & -0.0435 + 0.0065i  \\
-0.0007 - 0.0113i  & 0.0426 + 0.0024i  & 0.9990 + 0.0112i
\end{array}\right) \, ,
\label{CKMmatPNN}
\end{equation}  
   
\begin{equation}
\left|V_{CKM(24)}^{(+--)}\right|=\left(\begin{array}{ccc}
0.9756 & 0.2197 & 0.0034  \\                     
0.2194 & 0.9746 & 0.0440 \\
0.0113 & 0.0426  & 0.9990 
\end{array}\right) \, ,
\label{absCKMmatPNN}
\end{equation}  

\begin{equation}
\bar{\rho}_{(24)}^{(+--)} = 0.13\, , \,\,\,\,\, \bar{\eta}_{(24)}^{(+--)} = 0.31\, , 
\,\,\,\,\, J_{CP(24)}^{(+--)}=-2.9\times 10^{-5} \, .
\label{rhoetavalPNN}
\end{equation}

By looking at the four cases one can notice that all mass matrices have 
almost democratic structure with deviations from democracy for the up and down sector 
which depend on the different cases. In particular the situation with all 
components localized at the same orbifold fixed point $(+++)$ 
has both mass matrices very close to a DMM.  
The mass matrices, except the different Yukawa prefactors, are very similar.  
In this case a small top mass seems to be favored (Fig. \ref{Figure_3}).
For the configuration with the doublets localized at the zero orbifold
fixed point and both the up and down-singlets  at the other orbifold fixed point $(+--)$, 
the deviations from a pure democratic mass matrix are large for both mass matrices. 
Also in this case a small top mass seems to be favored (Fig. \ref{Figure_18}).
The situation is different in the other two cases where the up and down-type singlets 
are localized at different orbifold fixed points. 
In particular the case with the doublets and down-type singlets right components localized 
at the zero  fixed point and the up-type singlets at the other orbifold fixed point $(+-+)$ 
seems to be the one which allows a larger range for the top-quark mass values (Fig. \ref{Figure_13}).
In this case the deviation from a pure democratic mass matrix for the up sector 
is bigger than the one for the down sector.  
As we will show in the next section this is also the only case for 
which we were able to find  solutions for the 15 parameter version of the model. 
The forth case where the doublets and up-type singlets are at the same 
orbifold fixed point while the down-type singlets are at the other orbifold 
fixed point $(++-)$ gives for the top-quark mass a very narrow value-region around $178 GeV$ 
(Fig. \ref{Figure_8}). 
In this case the mass matrix for the up sector is very close to a pure democratic mass 
matrix while the deviation from it is larger for the down sector.
What is important to say here is that by looking at the four different cases, it seems  
that deviation from a DMM are bigger when left and right components are localized 
at different orbifold fixed points.   

\section{15 parameter version}

In this section we present the results for another particular choice of the 
model with 15 parameters, 
which correspond to all the Yukawa couplings with the same absolute 
value, $|F_{q,i}| = |F_{u,i}| = |F_{d,i}| = 1$. 
The family symmetry and left-right symmetry are now broken only 
through the parameters $\mu$'s  and phases $\theta$'s.  
The important point is that for the 15 parameter model choice we were able 
to find solutions only for the case corresponding to $f_q>0,\,f_u<0,\,f_d>0$ with the conditions 
that all $\mu$'s are bigger than one (\ref{compositesol}). 
In the other three cases we were not be able to find solutions if we decided to keep the constraints 
$\mu 's > 1$. 
The fact that we found solutions only for one of the four possible cases does not obviously exclude 
completely the existence of solutions for the other three cases, but we believe that we can at least 
conclude that the configuration   $f_q>0,\,f_u<0,\,f_d>0$ is favored respect to the others.     

In the following we give the solutions for the model's parameters and physics quantities as in the cases 
of 24 parameter version (see also Figs. \ref{Figure_21}-\ref{Figure_25})

\begin{itemize}
\item{$F_{q,i}=1 \,\, F_{u,i}=-1 \,\, F_{d,i}=1$}
\end{itemize}

\begin{equation}
g_{Yu(15)}^{(+-+)}\frac{v}{\sqrt{2}} = 60.69 \, , 
\,\,\, g_{Yd(15)}^{(+-+)}\frac{v}{\sqrt{2}} = 1.08 \, ,
\end{equation}
\begin{equation}
\m_{q1(15)}^{(+-+)} = 2.513 \, , \,\,\, \m_{q2(15)}^{(+-+)} =1.928 \, , \,\,\, 
\m_{q3(15)}^{(+-+)} = 1.993 \, ,
\end{equation}
\begin{equation}
\m_{u1(15)}^{(+-+)} = 1.177 \, , \,\,\, \m_{u2(15)}^{(+-+)} = 1.562 \, , \,\,\, 
\m_{u3(15)}^{(+-+)}= 1.152 \, ,
\end{equation}
\begin{equation}
\m_{d1(15)}^{(+-+)} = 4.969 \, , \,\,\, \m_{d2(15)}^{(+-+)} = 5.427 \, , \,\,\, 
\m_{d3(15)}^{(+-+)} = 1.022 \, .
\end{equation}
\begin{equation}
\phi_{u1(15)}^{(+-+)} =0.0153 \, , \,\,\, \phi_{u2(15)}^{(+-+)} = -0.0001 \, ,  
\,\,\,\phi_{d1(15)}^{(+-+)} =-0.0423 \, , \,\,\, \phi_{d2(15)}^{(+-+)} = -0.0279 \, ,
\end{equation}

\begin{equation}
M_{u(15)}^{(+-+)}= 60.69\,GeV\,\left(\begin{array}{ccc}
0.8968 & \,0.8619\,e^{i\,0.0153}   & 0.8986 \\           
0.9520 & 0.9267 & \,0.9532\,e^{-i\,0.0001}  \\
0.9469 & 0.9204 & 0.9481
\end{array}\right) \, ,
\label{matrixupPNP15}
\end{equation}  
\begin{equation}
m_{u(15)}^{(+-+)} = 0.0024 \,GeV\,,\,\,\,\, m_{c(15)}^{(+-+)}= 0.713\,GeV\,,
\,\,\,\, m_{t(15)}^{(+-+)}= 168.1\,GeV\, ,
\label{eigenupPNP15}
\end{equation}
\begin{equation}
M_{d(15)}^{(+-+)}= 1.08\, GeV \,\left(\begin{array}{ccc}
0.9086 & \,0.8874\,e^{-i\,0.0423}   & 0.9450 \\           
0.8414 & 0.8158   & \,0.9829\,e^{-i\,0.0279}  \\
0.8496 & 0.8245  & 0.9798
\end{array}\right) \, ,
\label{matrixdownPNP15}
\end{equation}  
\begin{equation}
m_{d(15)}^{(+-+)} =  0.0052\,GeV\,,\,\,\,\, m_{s(15)}^{(+-+)}= 0.084\,GeV\,,
\,\,\,\, m_{b(15)}^{(+-+)}= 2.9\,GeV\, .
\label{eigendownPNP15}
\end{equation}

\begin{equation}
U_{L(15)}^{u\dag (+-+)}=\left(\begin{array}{ccc}
-0.0336 - 0.0417i   & -0.6876 + 0.0393i   & 0.7231  \\           
0.8308 &-0.4195 + 0.0342i   &  -0.3622 - 0.0382i  \\
0.5540 - 0.0027i & 0.5904  & 0.5869
\end{array}\right) \, ,
\label{UudagaPNP15}
\end{equation}  

\begin{equation}
U_{L(15)}^{d(+-+)}=\left(\begin{array}{ccc}
-0.1221 - 0.1091i & 0.7910 & 0.5895 \\           
-0.6283 + 0.1216i & -0.5048 + 0.1047i  & 0.5698 + 0.0010i \\
0.7508 &-0.3108 - 0.1091i  &   0.5725 + 0.0074i
\end{array}\right) \, ,
\label{UdPNP15}
\end{equation}

\begin{equation}
V_{CKM(15)}^{(+-+)}=\left(\begin{array}{ccc}
0.9696 - 0.0995i &0.0917 - 0.2038i  & 0.0024 + 0.0025i \\           
-0.1140 - 0.1919i & 0.9737 - 0.0098i  & 0.0436 - 0.0055i \\
0.0018 + 0.0116i & -0.0422 - 0.0044i  & 0.9990 + 0.0033i
\end{array}\right) \, ,
\label{CKMmatPNP15}
\end{equation}  
   
\begin{equation}
\left|V_{CKM(15)}^{(+-+)}\right|=\left(\begin{array}{ccc}
0.9747& 0.2235  & 0.0034  \\           
0.2232 & 0.9738 & 0.0439 \\
0.0118 & 0.0424 & 0.9990 
\end{array}\right) \, ,
\label{absCKMmatPNP15}
\end{equation}  

\begin{equation}
\bar{\rho}^{(+-+)}_{(15)} = 0.16\, , \,\,\,\,\, \bar{\eta}^{(+-+)}_{(15)} = 0.30\, , 
\,\,\,\,\, J_{CP(15)}^{(+-+)}=-2.9\times 10^{-5} \, .
\label{rhoetavalPNP15}
\end{equation}

As it can be seen in the numerical example given above, also in the case of the 15 
parameter version, as for all the  24 parameter cases, 
both mass matrices for the up and down sector are almost democratic. 
What has to be also noticed is that the 15 parameter case favors a small top mass 
(Fig. \ref{Figure_23}),  
on the contrary to the 24 parameter corresponding case $(+-+)$  which gives a much larger range for 
the top-quark mass.   

%
%

\section{Epilogue}
We suggest that using one extra dimension compactified on an $S_1/Z_2$ orbifold 
one is able to produce an almost 
democratic mass matrix and obtain the right mass spectrum and right CKM matrix. 
In the model presented the zero modes  
are localized only at the orbifold fixed points and different profiles for the zero 
mode wave functions are allowed. 
We show that in the case of the 24 parameter version of the model, for all four 
possible scenarios to localize the left and right handed components of quarks 
at one or the other orbifold fixed point, we were able to fit the mass spectrum and 
CKM matrix. On the other hand in the case of the 15 parameter version of the model, 
which corresponds to having the universal absolute value of 
the Yukawa couplings with the background scalar field for the different fermion 
families, we were able to reproduce the right 
mass spectrum and right CKM only in the case with the doublets and down-type singlets
localized at one orbifold fixed point 
and the up-type singlets at the other orbifold fixed point. 
Finally we just also explain how the existence of a sixth dimension could account for 
the different Yukawa couplings for the up and down sectors.

\acknowledgments{We would like to thank Prof. P.Q. Hung and Dr. M. Seco for valuable discussions, 
and the University of Virginia High Energy Theory Group for supporting our work.}

\newpage
\begin{table}[!ht]
\caption{Central values and uncertainties for the masses of the 6 quarks 
evaluated at $M_Z$, for 
the two ratios $m_u/m_d$ and $m_s/m_d$, for the absolute values of the CKM matrix elements and 
the CP parameters $\bar{\rho}, \bar{\eta}$}
\begin{center}
\begin{ruledtabular}
\begin{tabular}{ccc}
$x_i$ & $<x_i>$ & $|x^{max}_i-x^{min}_i|/2$ \\ \hline \\
$m_u$ & $2.33 \times 10^{-3}$ & $0.45 \times 10^{-3}$ \\ 
$m_c$ & $0.685 $ & $0.061$ \\ 
$m_t$ & $181$ & $13$ \\ 
$m_d$ & $4.69 \times 10^{-3}$ & $0.66 \times 10^{-3}$ \\ 
$m_s$ & $0.0934 $ & $0.0130$ \\ 
$m_b$ & $3.00$ & $0.11$ \\ 
$m_u/m_d$ & $0.497$ & $0.119$ \\ 
$m_s/m_d$ & $19.9$ & $3.9$ \\ 
$|V_{ud}|$ & $0.97485$  & $0.00075$ \\ 
$|V_{us}|$ & $0.2225$  & $0.0035$ \\ 
$|V_{ub}|$ & $0.00365$  & $0.0115$ \\ 
$|V_{cd}|$ & $0.2225$  & $0.0035$ \\ 
$|V_{cs}|$ & $0.9740$  & $0.0008$ \\ 
$|V_{cb}|$ & $0.041$  & $0.003$ \\ 
$|V_{td}|$ & $0.009$  & $0.005$ \\ 
$|V_{ts}|$ & $0.0405$  & $0.0035$ \\ 
$|V_{tb}|$ & $0.99915$  & $0.00015$ \\ 
$\bar{\rho}$ & $0.22$ & $0.10$ \\ 
$\bar{\eta}$ & $0.35$ & $0.05$  \\ 
\end{tabular}
\end{ruledtabular}
\end{center}
\end{table}


\newpage
\begin{figure}[!ht]
\caption{Summary of the 24 parameter space corresponding to $f_{qi}>0 \,\, f_{ui}>0 \,\, f_{di}>0$. \label{Figure_1}}
\end{figure}

\begin{figure}[!ht]
\caption{Profile of the VEV's and of the wave functions for left and right components corresponding to 
$f_{qi}>0 \,\, f_{ui}>0 \,\, f_{di}>0$ for the 24 parameter space. 
\label{Figure_2}}
\end{figure}

\begin{figure}[!ht]
\caption{Solutions for the 6 quark masses corresponding to 
$f_{qi}>0 \,\, f_{ui}>0 \,\, f_{di}>0$ for the 24 parameter space. 
The masses in $GeV$ are evaluated at the $M_Z$ scale. 
The range for each mass is given by the edges of the corresponding window. 
\label{Figure_3}}
\end{figure}

\begin{figure}[!ht]
\caption{Solutions for the absolute values of the CKM matrix elements corresponding 
to $f_{qi}>0 \,\, f_{ui}>0 \,\, f_{di}>0$ for the 24 parameter space.
The range for each element is given by the edges of the window. 
\label{Figure_4}}
\end{figure}

\begin{figure}[!ht]
\caption{Solutions for $\bar{\rho}$ and $\bar{\eta}$ 
corresponding to  $f_{qi}>0 \,\, f_{ui}>0 \,\, f_{di}>0$ for the 24 parameter space. 
The delimited area is the allowed region in the 
$\bar{\rho},\bar{\eta}$ plane. \label{Figure_5}}
\end{figure}

\begin{figure}[!ht]
\caption{Summary of the 24 parameter space corresponding to $f_{qi}>0 \,\, f_{ui}>0 \,\, f_{di}<0$. \label{Figure_6}}
\end{figure}

\begin{figure}[h!]
\caption{Profile of the VEV's and of the wave functions for left and right components corresponding to 
$f_{qi}>0 \,\, f_{ui}>0 \,\, f_{di}<0$ for the 24 parameter case. 
\label{Figure_7}}
\end{figure}

\begin{figure}[!ht]
\caption{Solutions for the 6 quark masses corresponding to 
$f_{qi}>0 \,\, f_{ui}>0 \,\, f_{di}<0$ for the 24 parameter case. 
The masses in $GeV$ are evaluated at the $M_Z$ scale. 
The range for each mass is given by the edges of the corresponding window. 
\label{Figure_8}}
\end{figure}

\begin{figure}[!ht]
\caption{Solutions for the absolute values of the CKM matrix elements corresponding 
to $f_{qi}>0 \,\, f_{ui}>0 \,\, f_{di}<0$ for the 24 parameter case.
The range for each element is given by the edges of the window. 
\label{Figure_9}}
\end{figure}

\begin{figure}[!ht]
\caption{Solutions for $\bar{\rho}$ and $\bar{\eta}$ corresponding to
$f_{qi}>0 \,\, f_{ui}>0 \,\, f_{di}<0$ for the 24 parameter case. 
The delimited area is the allowed region in the 
$\bar{\rho},\bar{\eta}$ plane. \label{Figure_10}}
\end{figure}

\begin{figure}[!ht]
\caption{Summary of the 24 parameter space corresponding to $f_{qi}>0 \,\, f_{ui}<0 \,\, f_{di}>0$. \label{Figure_11}}
\end{figure}

\begin{figure}[!ht]
\caption{Profile of the VEV's and of the wave functions for left and right components corresponding to 
$f_{qi}>0 \,\, f_{ui}<0 \,\, f_{di}>0$. \label{Figure_12}}
\end{figure}

\begin{figure}[!ht]
\caption{Solutions for the 6 quark masses corresponding to 
$f_{qi}>0 \,\, f_{ui}<0 \,\, f_{di}>0$ for the 24 parameter case. 
The masses in $GeV$ are evaluated at the $M_Z$ scale. 
The range for each mass is given by the edges of the corresponding window. 
\label{Figure_13}}
\end{figure}

\begin{figure}[!ht]
\caption{Solutions for the absolute values of the CKM matrix elements corresponding 
to $f_{qi}>0 \,\, f_{ui}<0 \,\, f_{di}>0$ for the 24 parameter case.
The range for each element is given by the edges of the window. 
\label{Figure_14}}
\end{figure}

\begin{figure}[!ht]
\caption{Solutions for $\bar{\rho}$ and $\bar{\eta}$ corresponding to 
$f_{qi}>0 \,\, f_{ui}<0 \,\, f_{di}>0$ 
for the 24 parameter case. 
The delimited area is the allowed region in the 
$\bar{\rho},\bar{\eta}$ plane. \label{Figure_15}}
\end{figure}

\begin{figure}[!ht]
\caption{Summary of the 24 parameter space corresponding to $f_{qi}>0 \,\, f_{ui}<0 \,\, f_{di}<0$. \label{Figure_16}}
\end{figure}

\begin{figure}[!ht]
\caption{Profile of the VEV's and of the wave functions for left and right components corresponding to 
$f_{qi}>0 \,\, f_{ui}<0 \,\, f_{di}<0$ for the 24 parameter case. 
\label{Figure_17}}
\end{figure}

\begin{figure}[!ht]
\caption{Solutions for the 6 quark masses corresponding to $f_{qi}>0 \,\, f_{ui}<0 \,\, f_{di}<0$ 
for the 24 parameter case. 
The masses in $GeV$ are evaluated at the $M_Z$ scale. 
The range for each mass is given by the edges of the corresponding window. 
\label{Figure_18}}
\end{figure}

\begin{figure}[!ht]
\caption{Solutions for the absolute values of the CKM matrix elements corresponding 
to $f_{qi}>0 \,\, f_{ui}<0 \,\, f_{di}<0$ for the 24 parameter case.
The range for each element is given by the edges of the window. 
\label{Figure_19}}
\end{figure}

\begin{figure}[!ht]
\caption{Solutions for $\bar{\rho}$ and $\bar{\eta}$ corresponding to 
$f_{qi}>0 \,\, f_{ui}<0 \,\, f_{di}<0$ for the 24 parameter case. 
The delimited area is the allowed region in the 
$\bar{\rho},\bar{\eta}$ plane. \label{Figure_20}}
\end{figure}

\begin{figure}[!ht]
\caption{Summary of the 15 parameter space corresponding to $f_{qi}>0 \,\, f_{ui}<0 \,\, f_{di}<0$. \label{Figure_21}}
\end{figure}

\begin{figure}[!ht]
\caption{Profile of the VEV's and of the wave functions for left and right components corresponding to 
$f_{qi}>0 \,\, f_{ui}<0 \,\, f_{di}<0$ for the 15 parameter case. 
\label{Figure_22}}
\end{figure}

\begin{figure}[!ht]
\caption{Solutions for the 6 quark masses corresponding to $f_{qi}>0 \,\, f_{ui}<0 \,\, f_{di}<0$ 
for the 15 parameter case. 
The masses in $GeV$ are evaluated at the $M_Z$ scale. 
The range for each mass is given by the edges of the corresponding window. 
\label{Figure_23}}
\end{figure}

\begin{figure}[!ht]
\caption{Solutions for the absolute values of the CKM matrix elements corresponding 
to $f_{qi}>0 \,\, f_{ui}<0 \,\, f_{di}<0$ for the 15 parameter case.
The range for each element is given by the edges of the window. 
\label{Figure_24}}
\end{figure}

\begin{figure}[!ht]
\caption{Solutions for $\bar{\rho}$ and $\bar{\eta}$ corresponding to 
$f_{qi}>0 \,\, f_{ui}<0 \,\, f_{di}<0$ for the 15 parameter case. 
The delimited area is the allowed region in the 
$\bar{\rho},\bar{\eta}$ plane. \label{Figure_25}}
\end{figure}

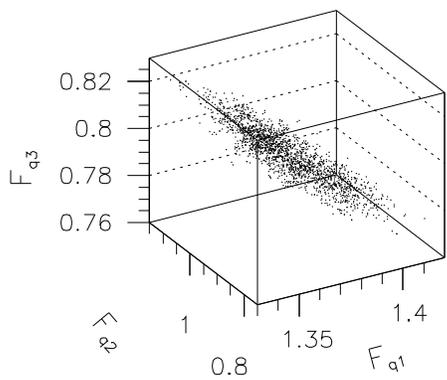
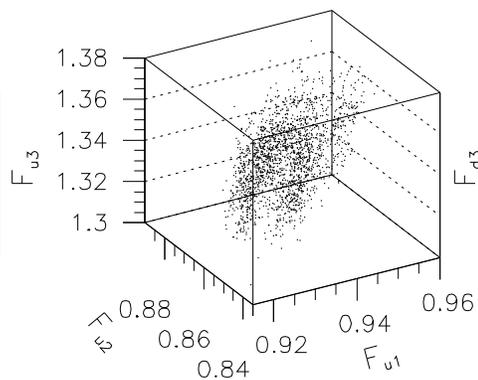
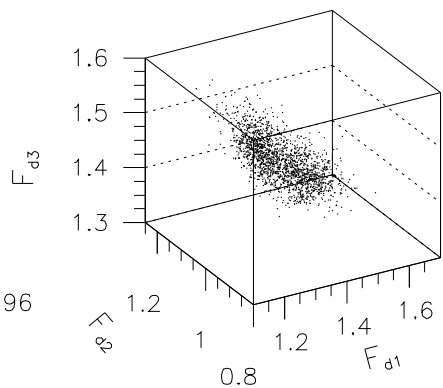
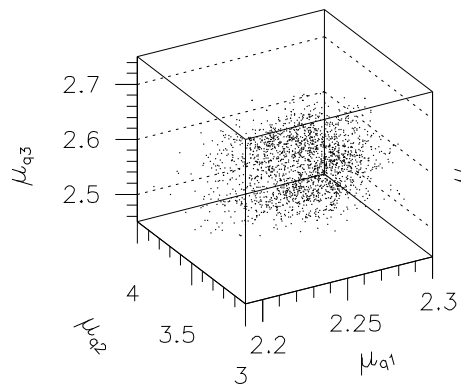
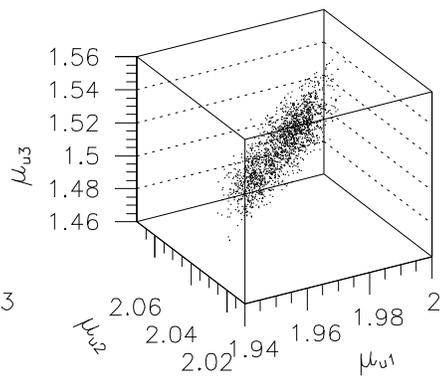
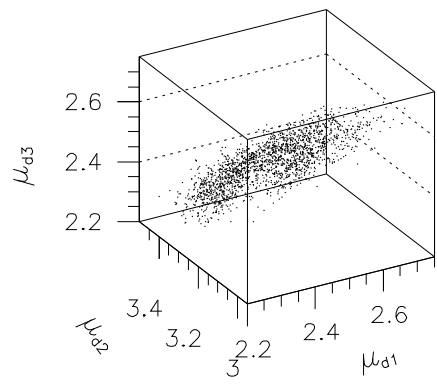
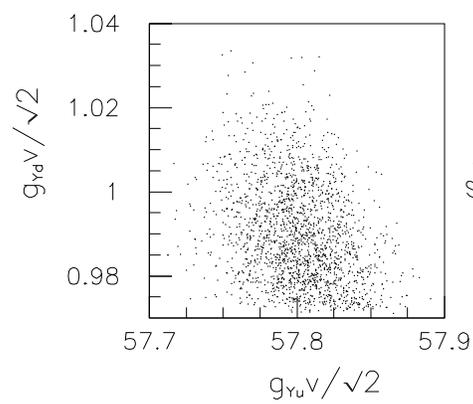
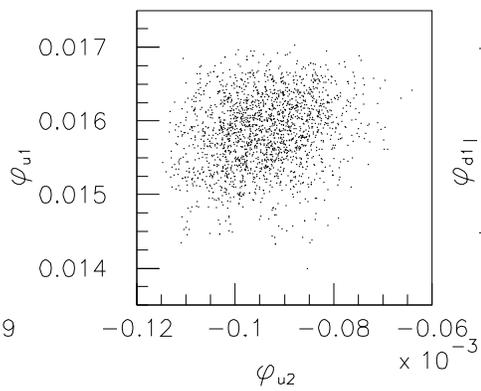
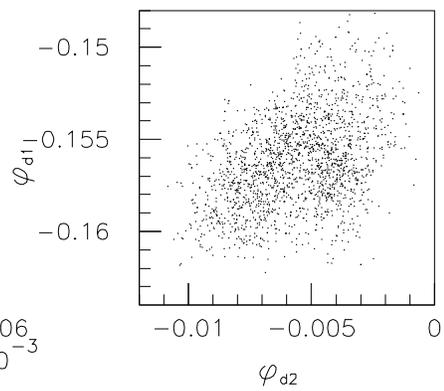
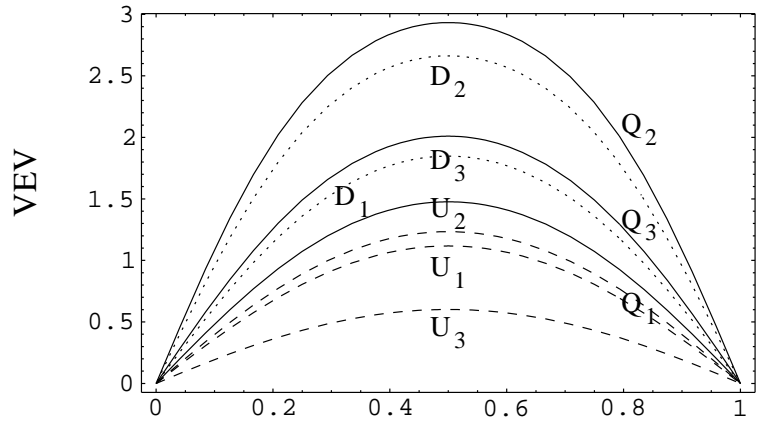
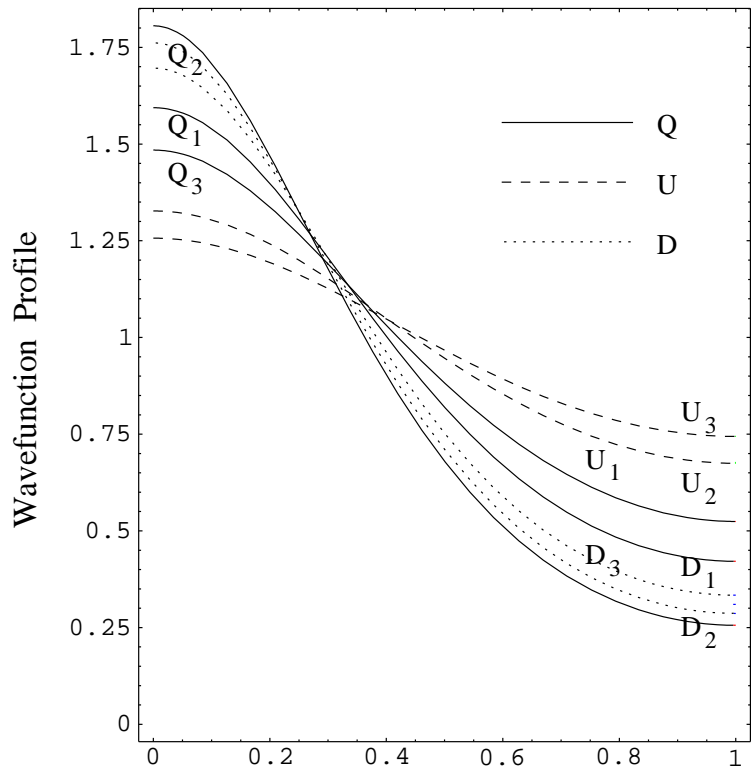
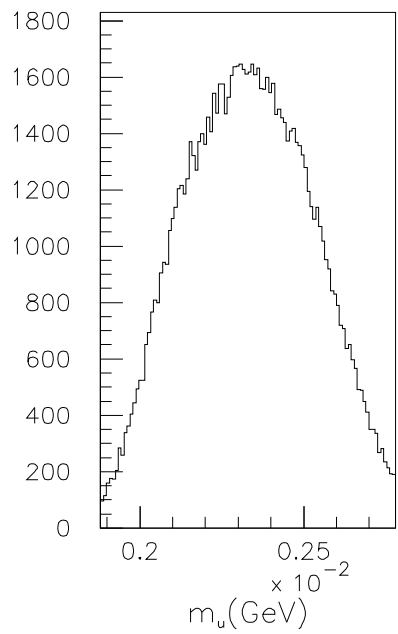
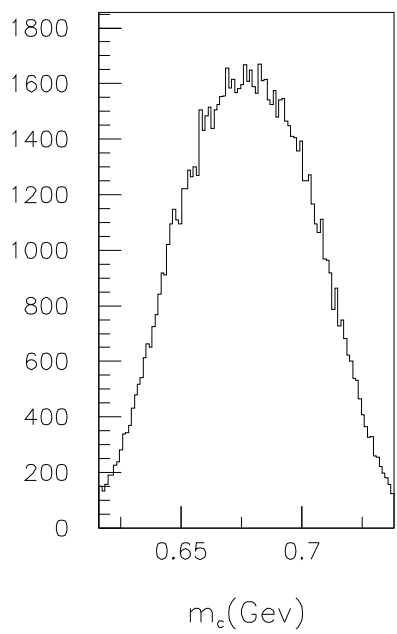
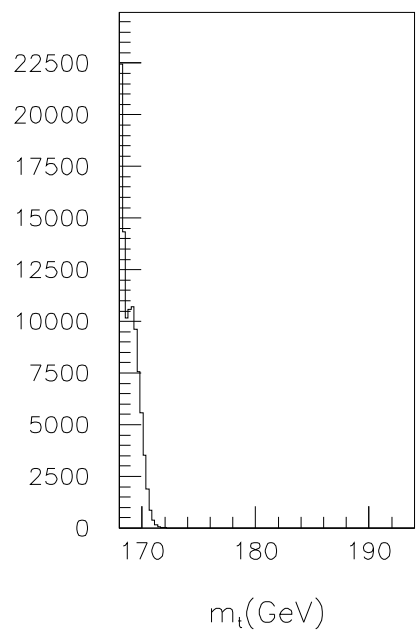
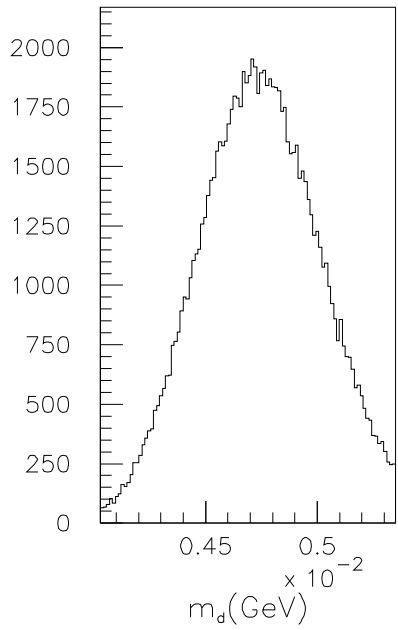
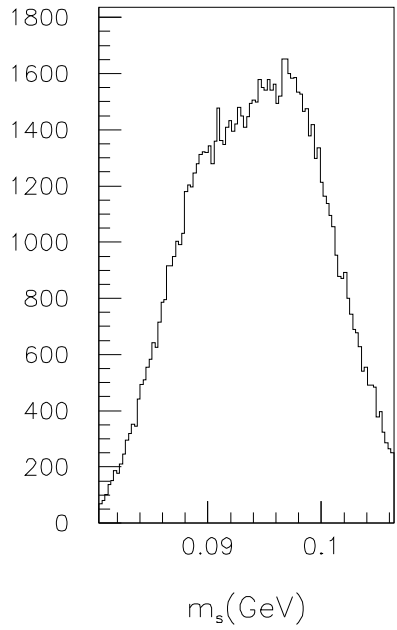
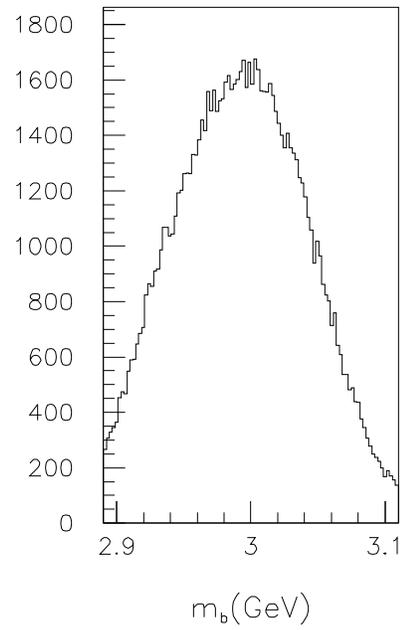
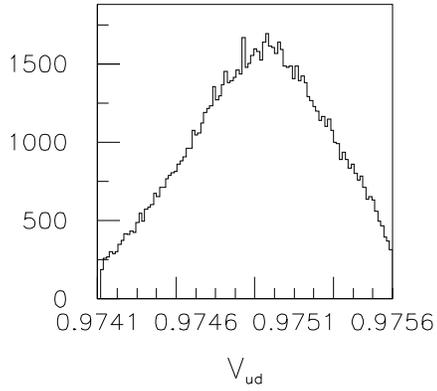
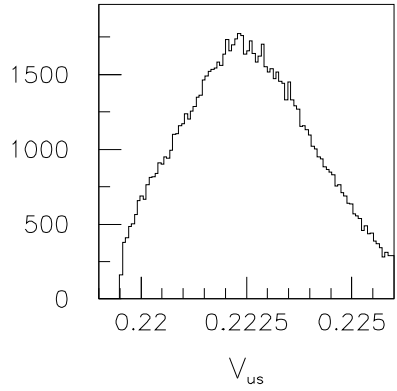
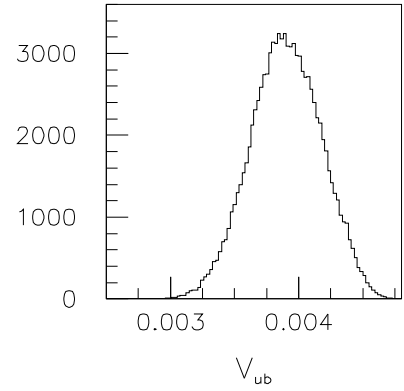
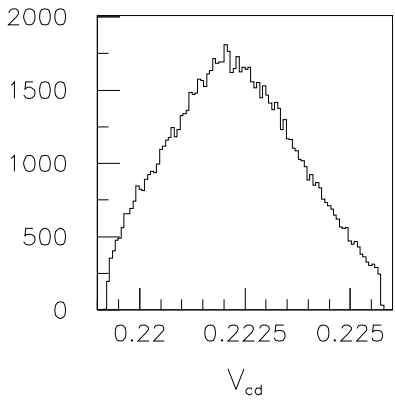
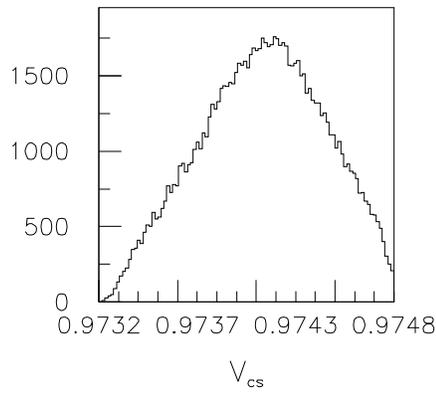
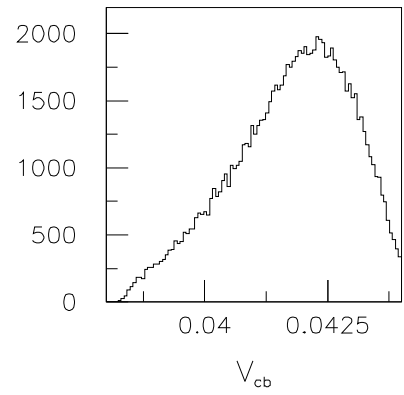
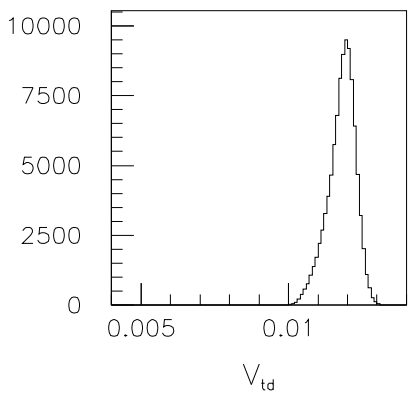
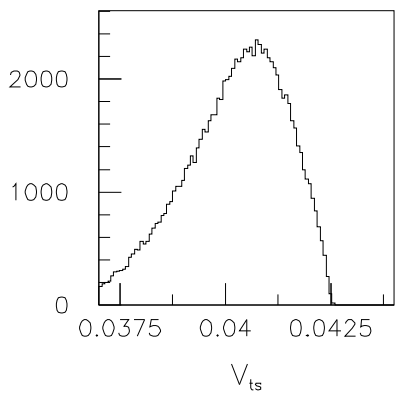
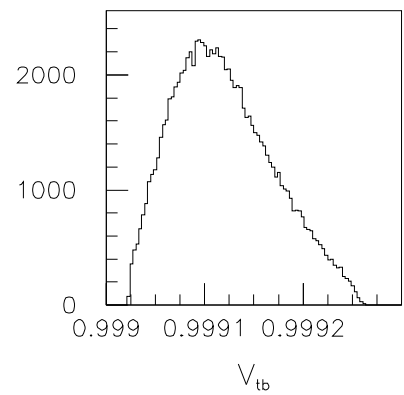
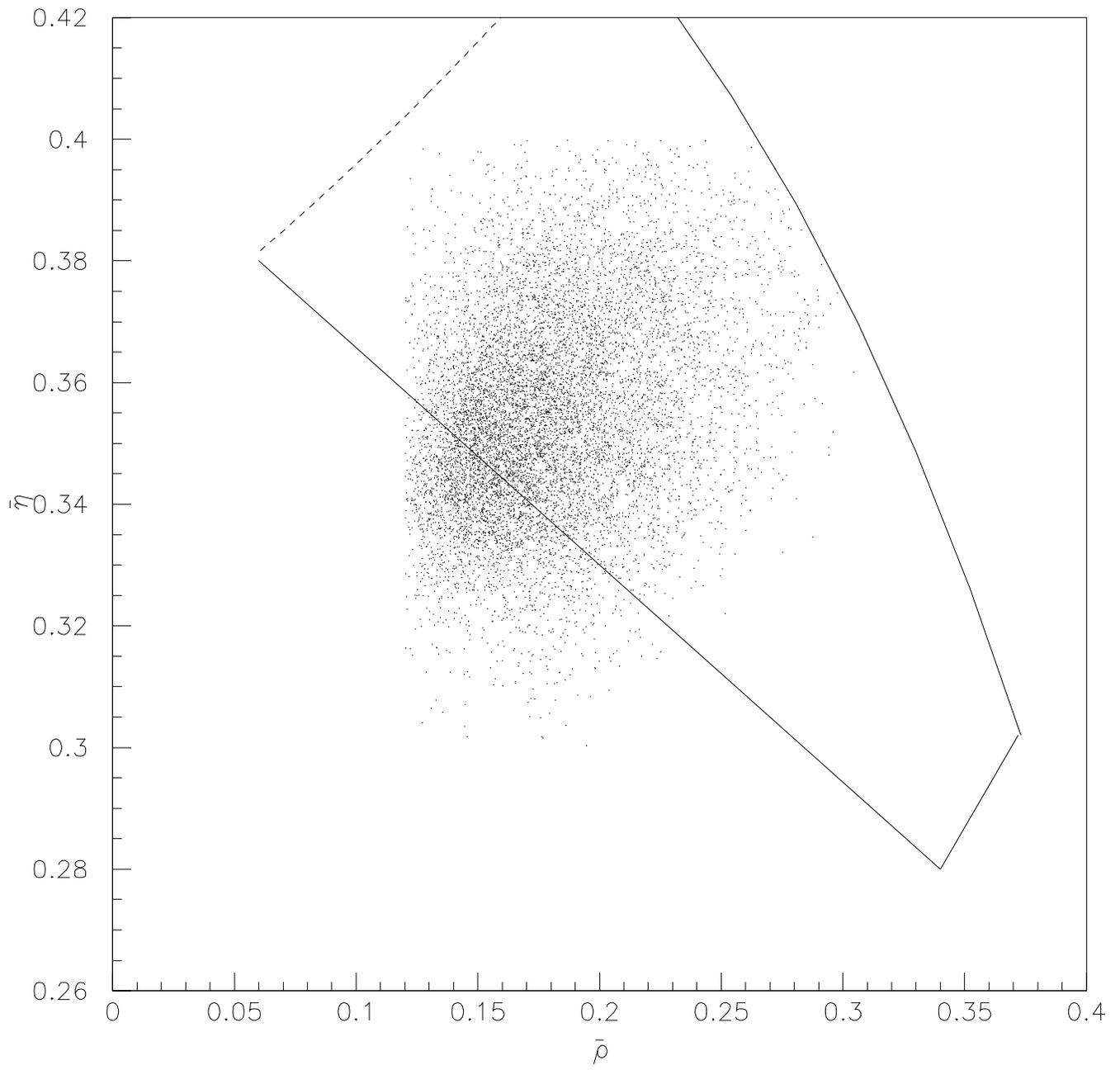
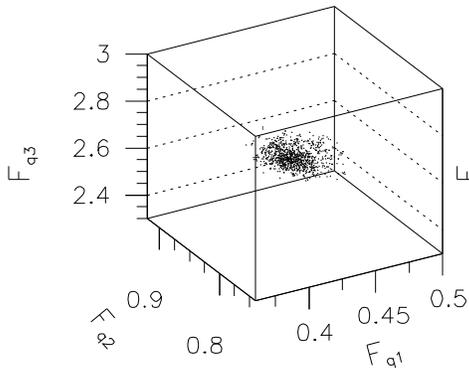
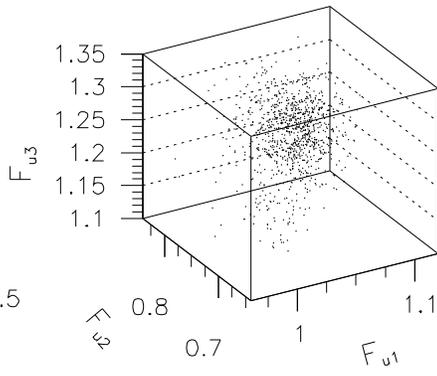
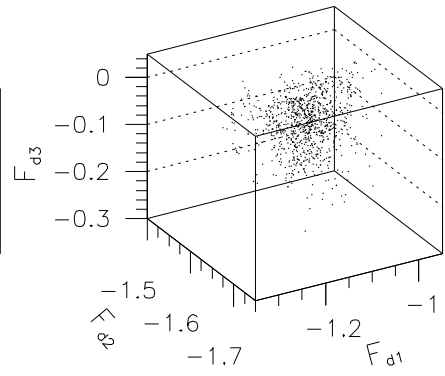
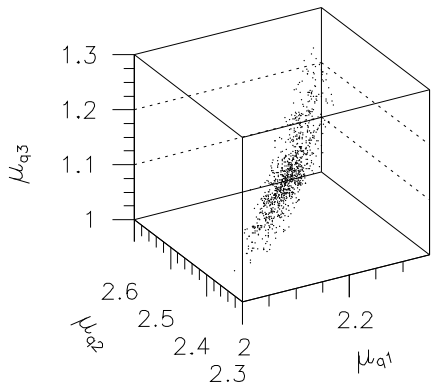
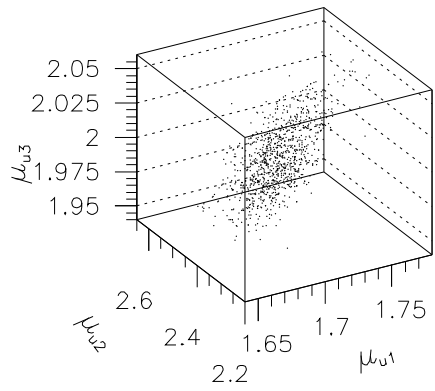
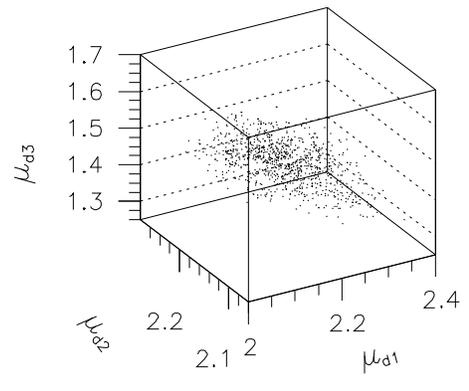
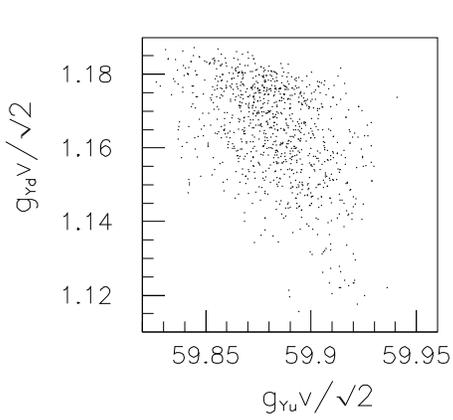
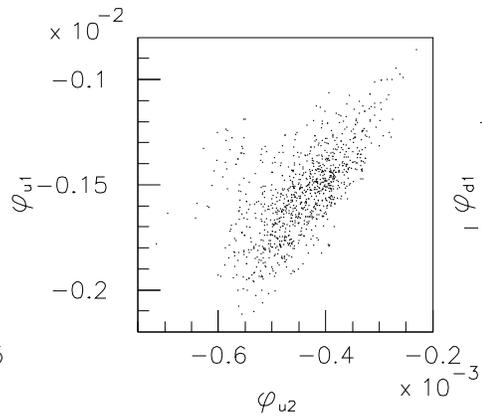
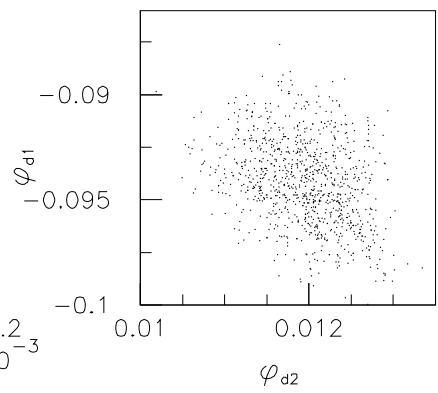
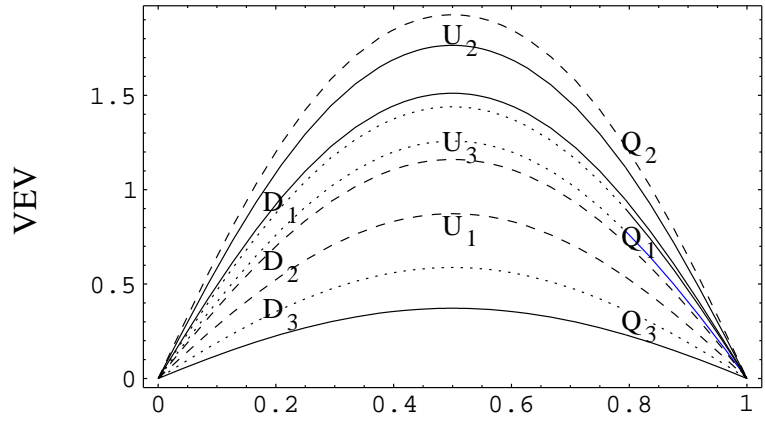
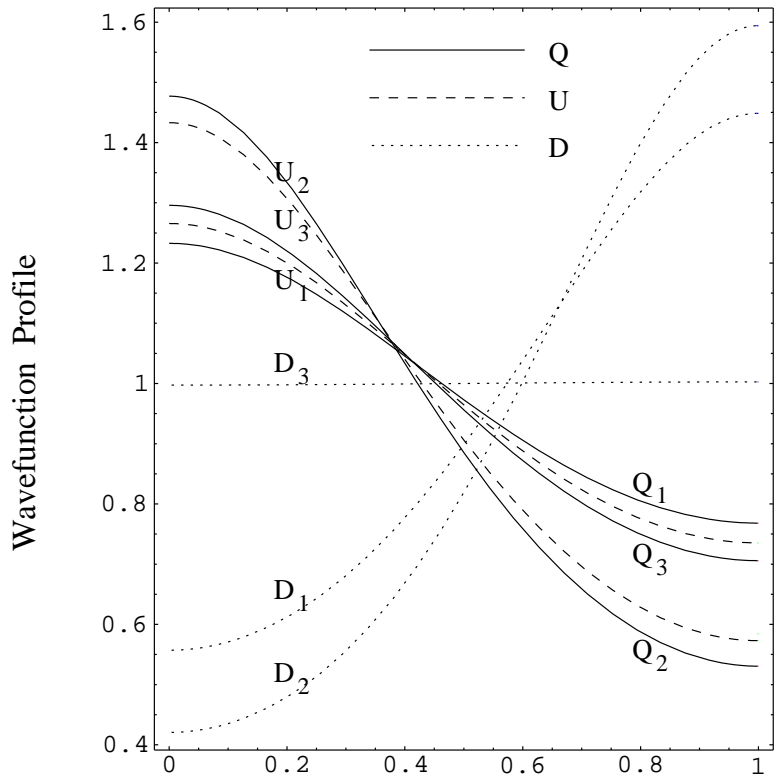
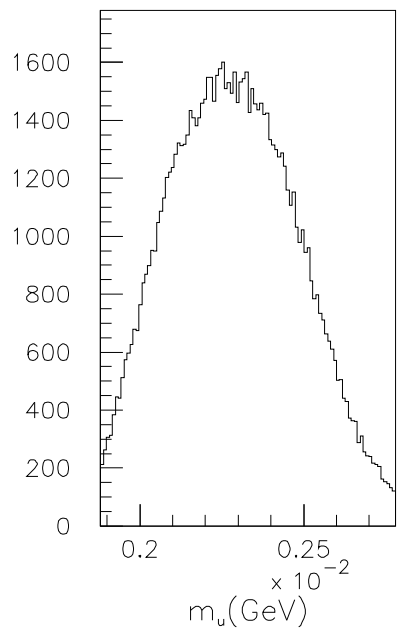
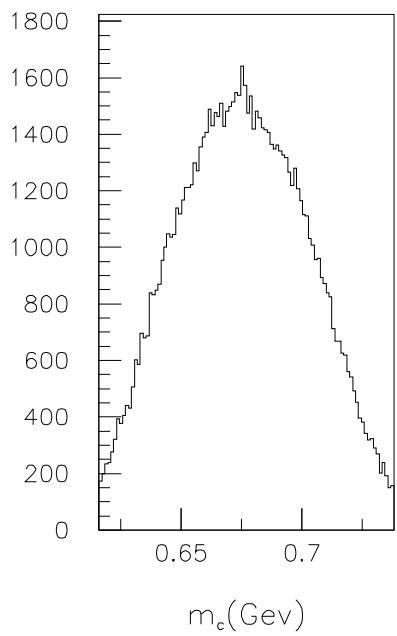
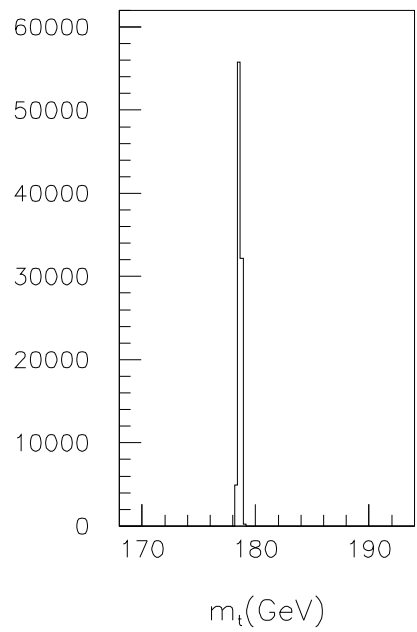
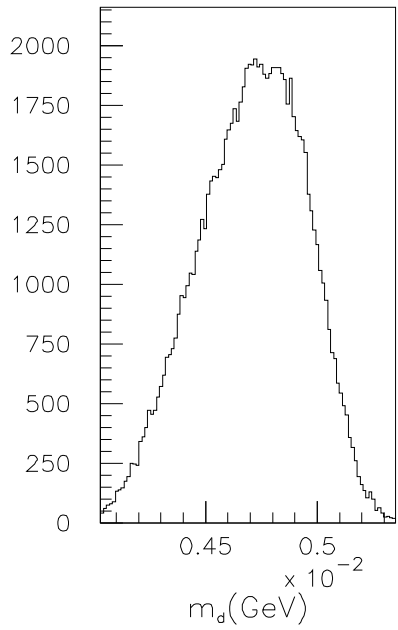
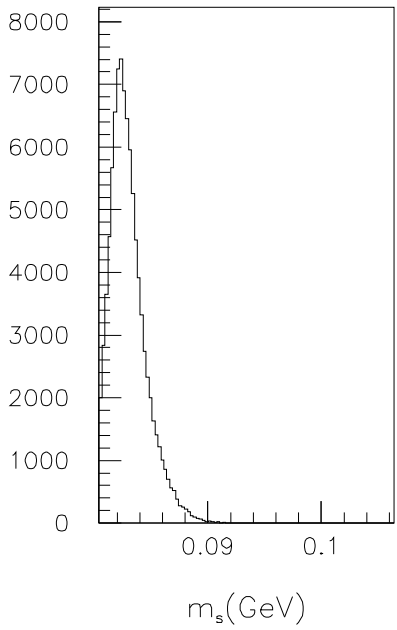
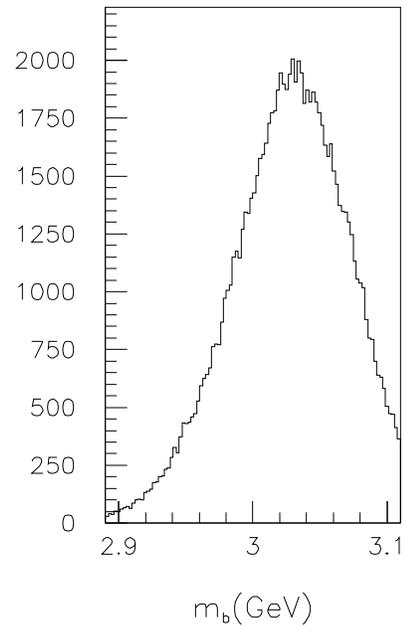
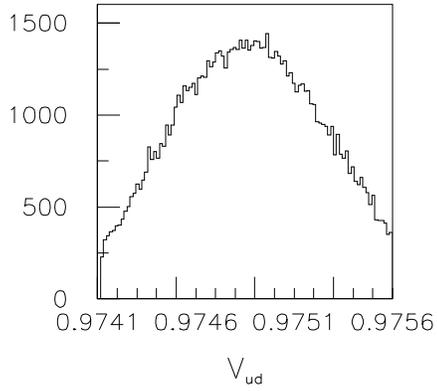
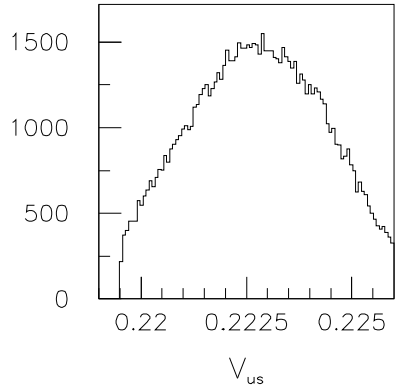
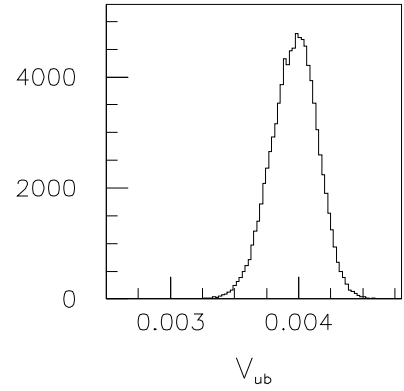
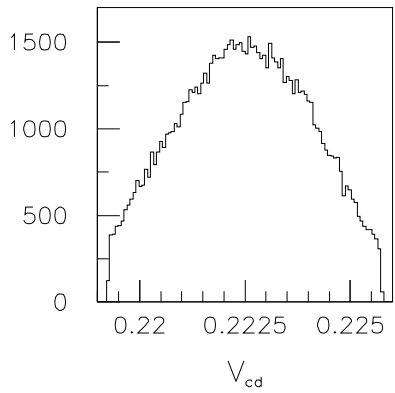
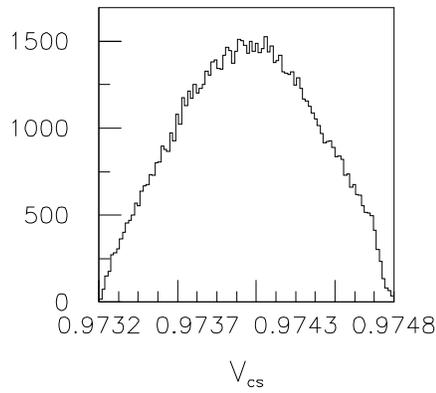
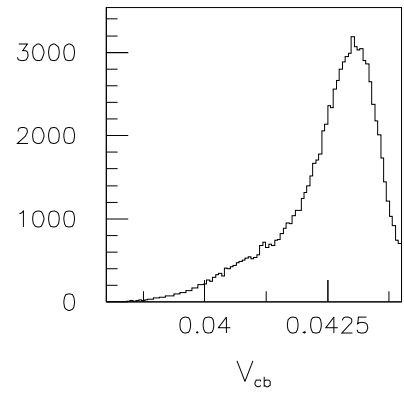
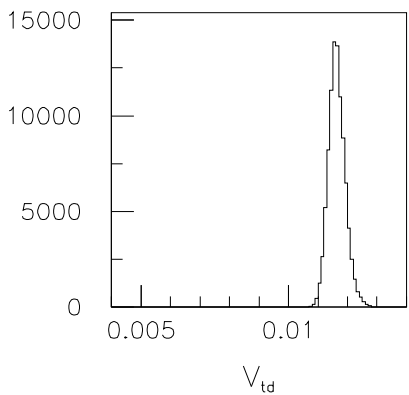
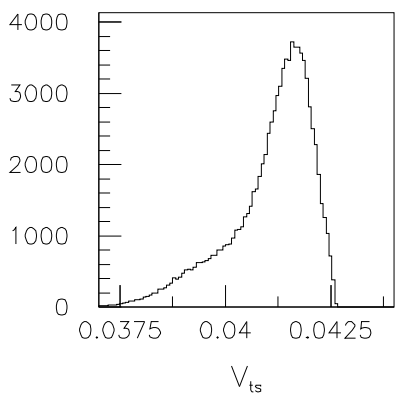
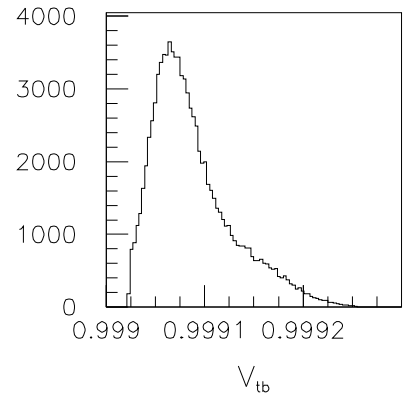
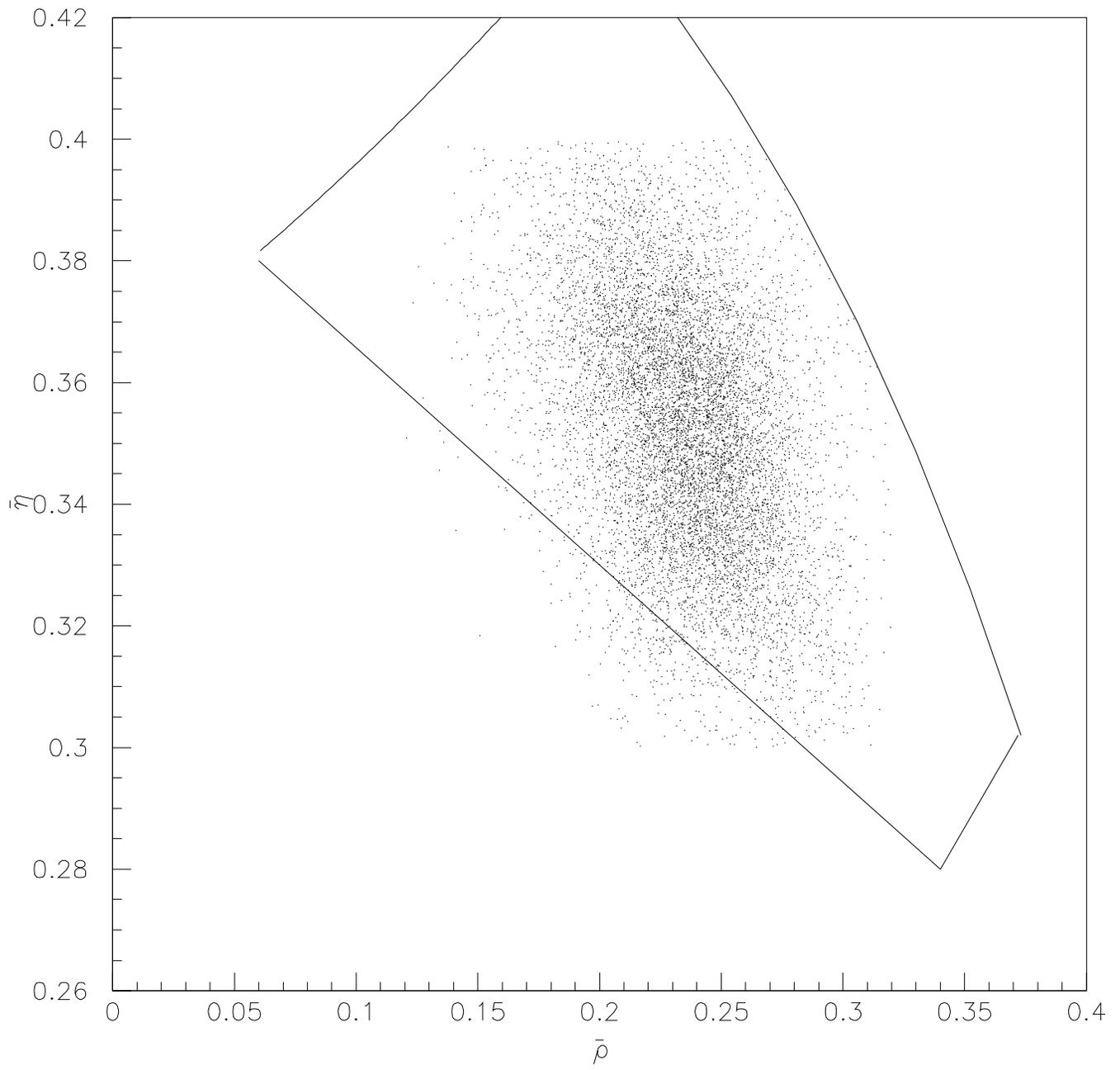
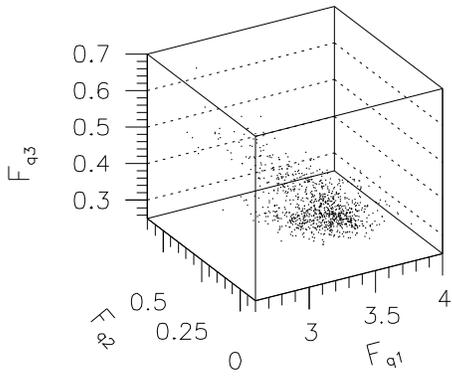
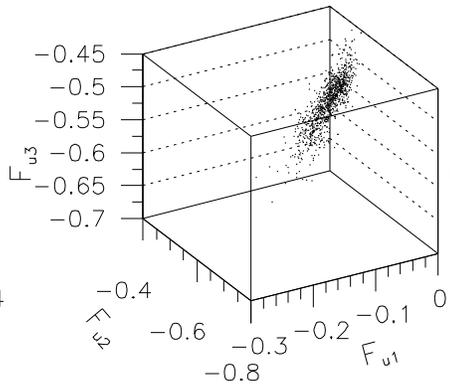
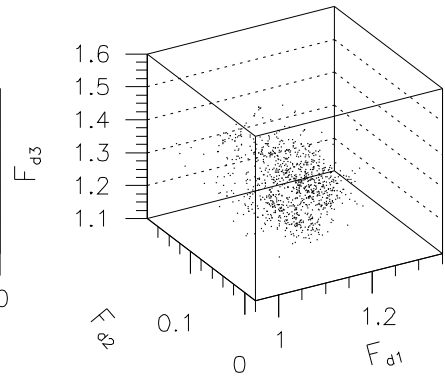
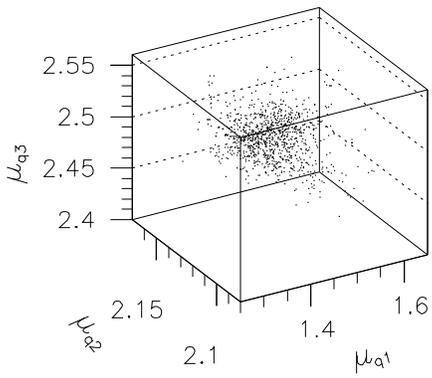
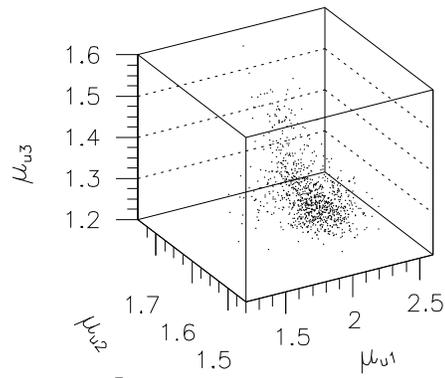
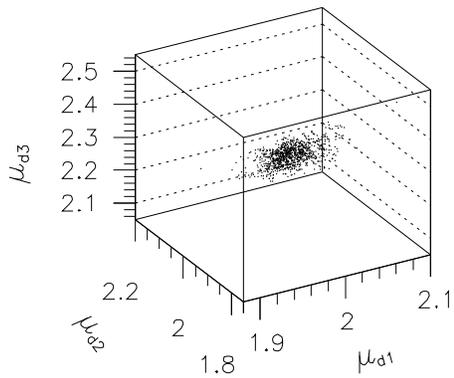
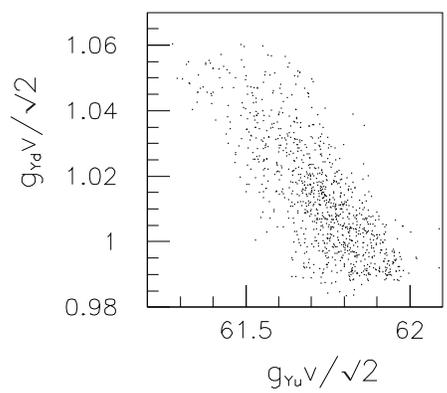
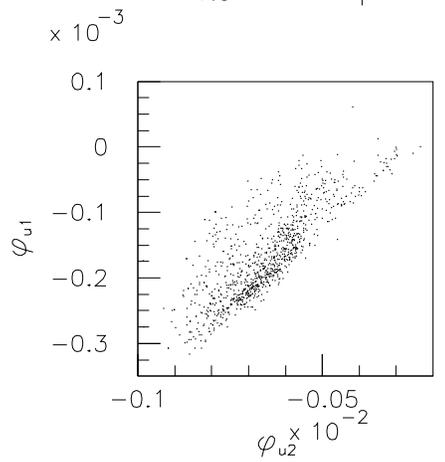
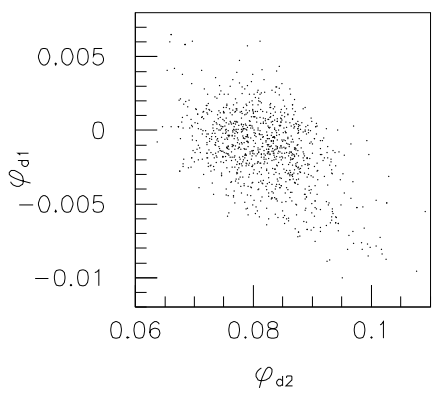
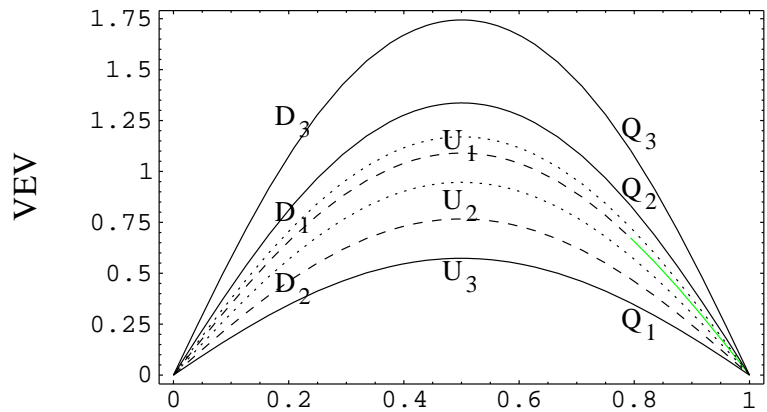
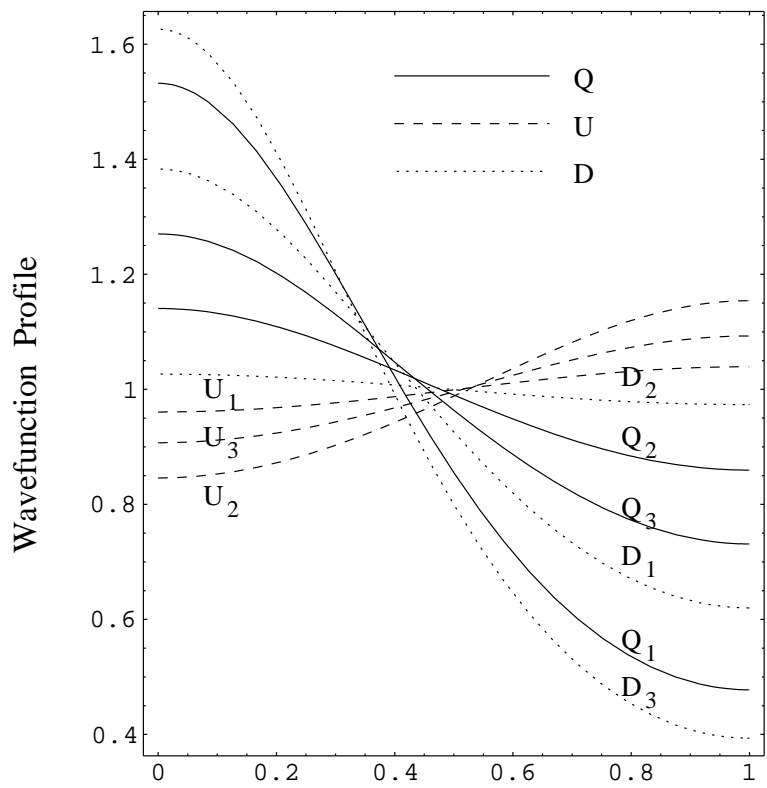
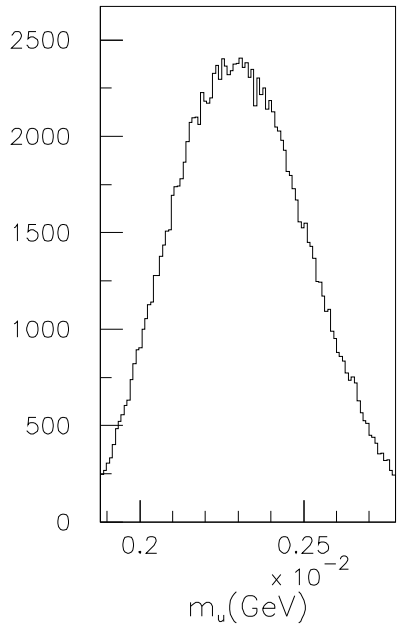
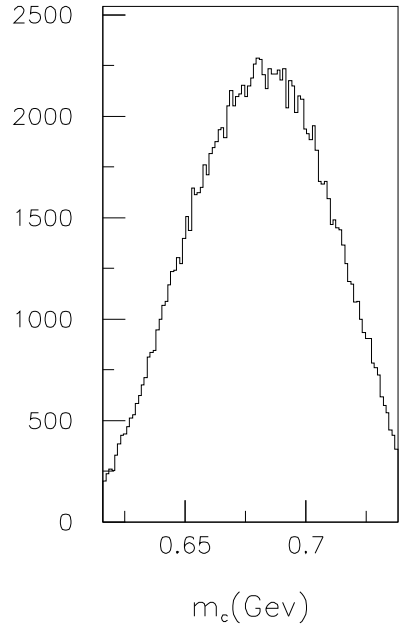
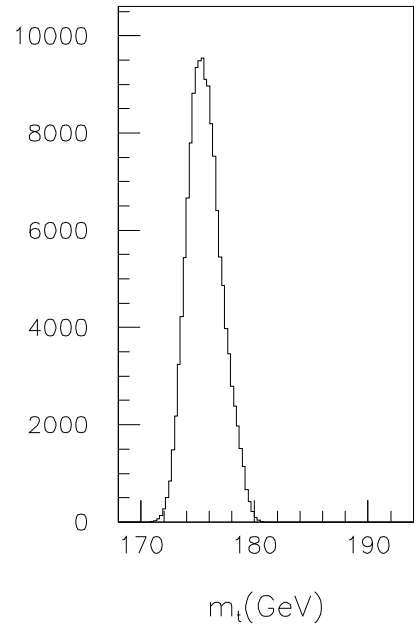
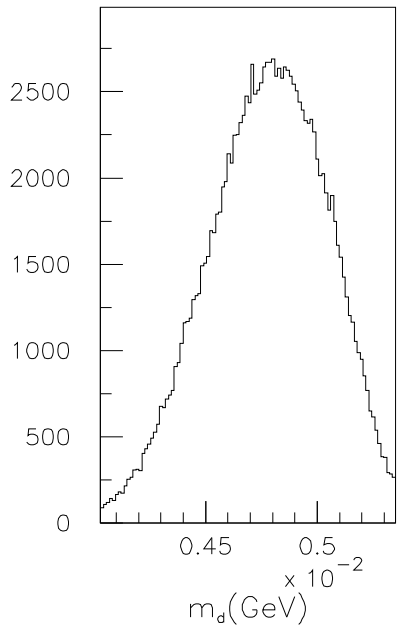
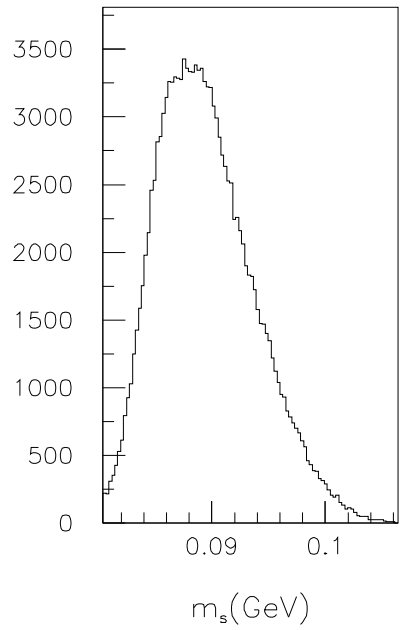
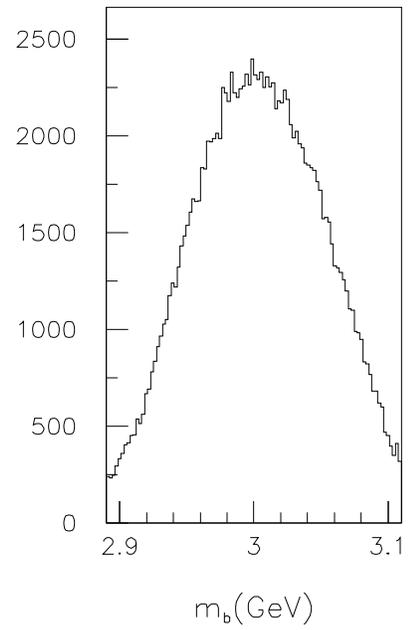
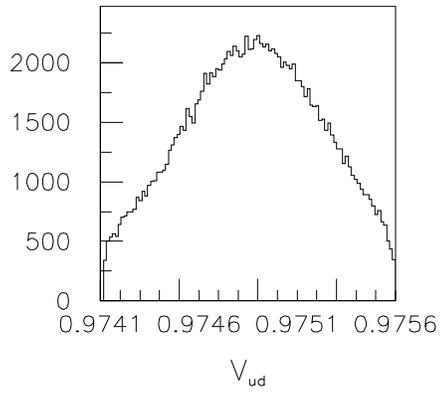
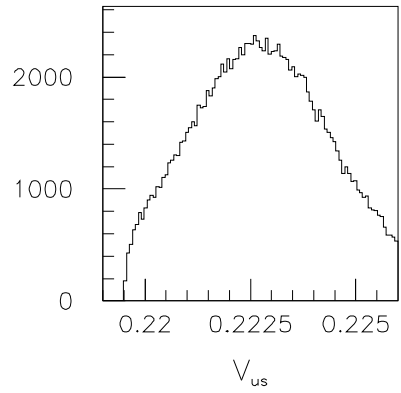
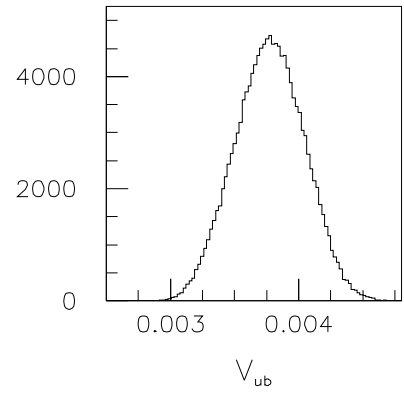
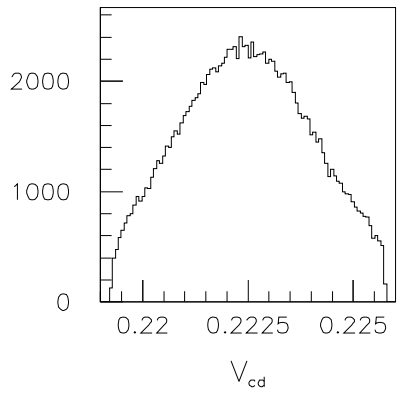
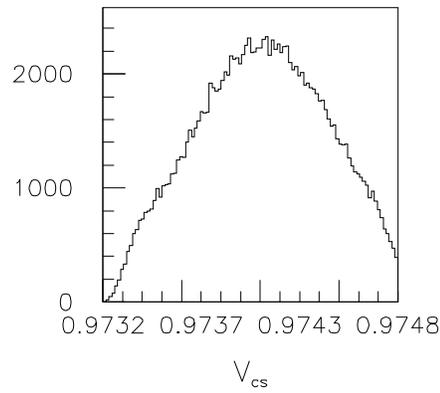
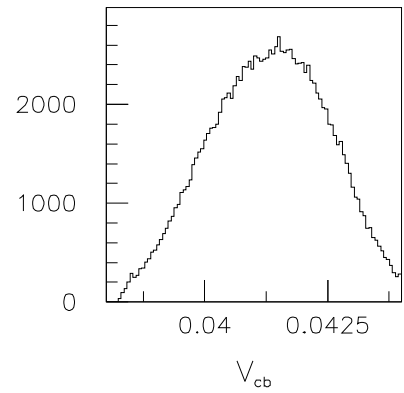
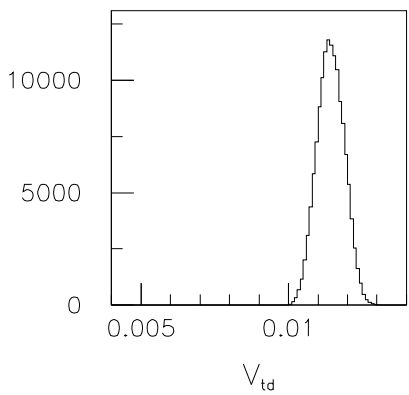
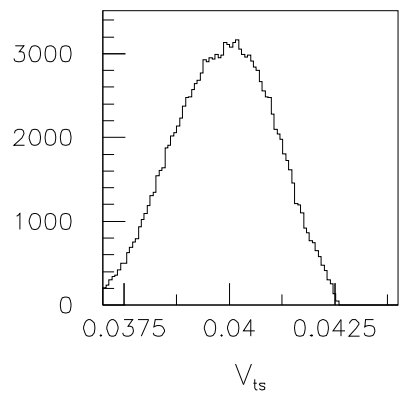
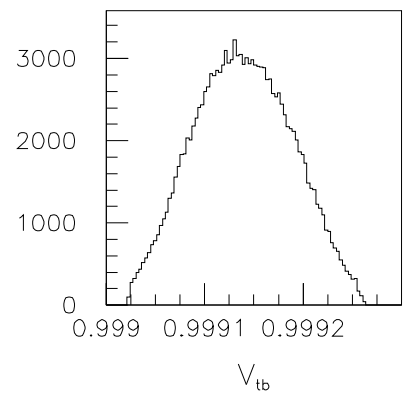
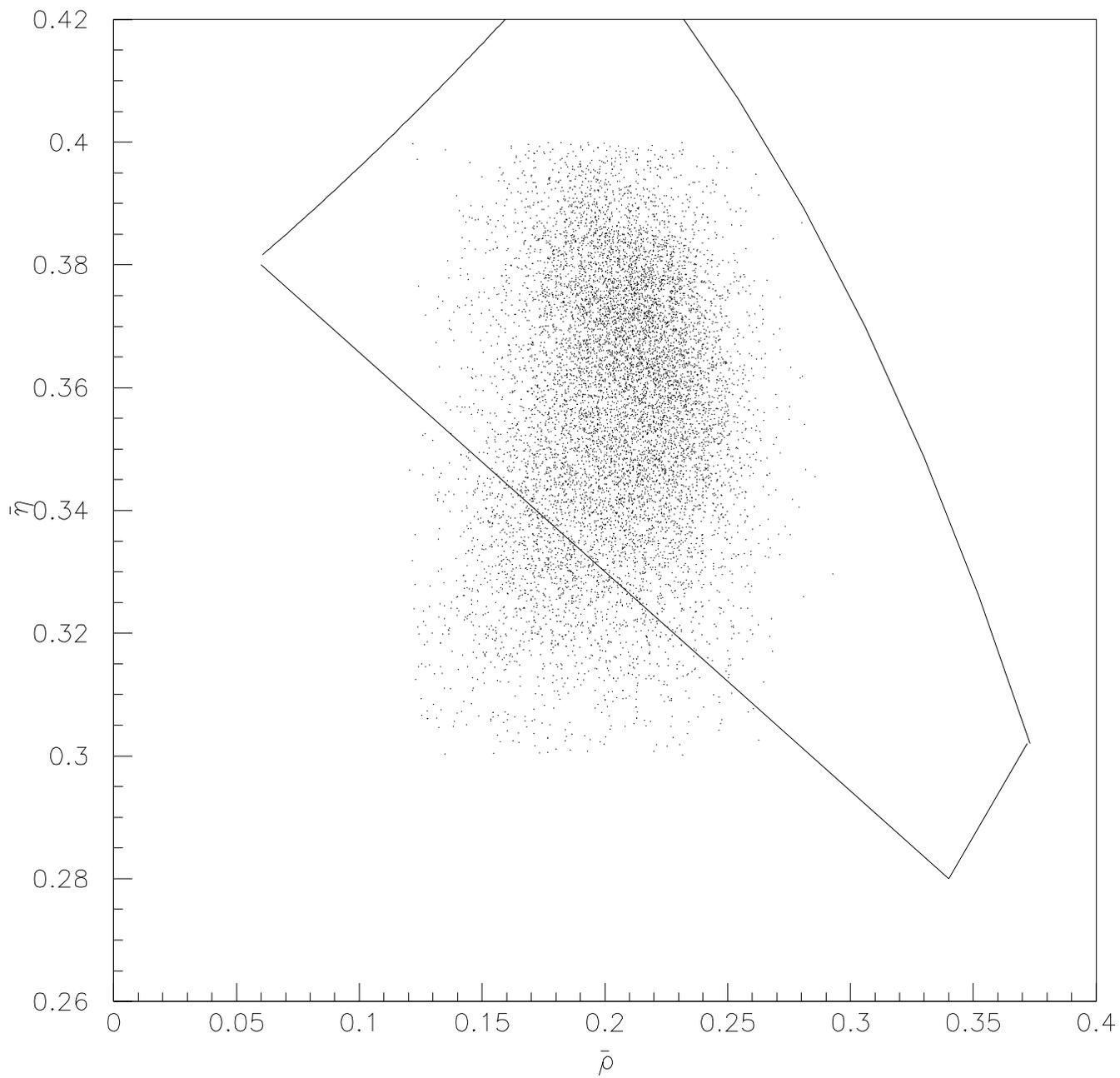
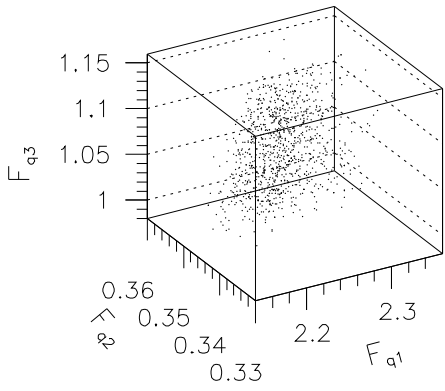
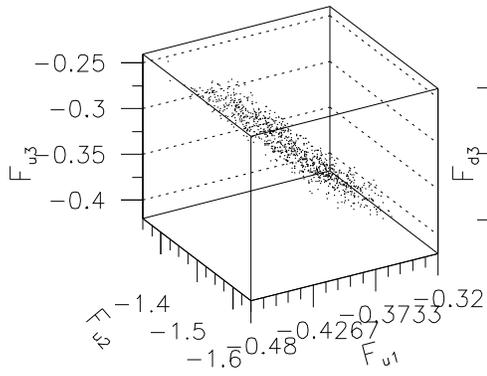
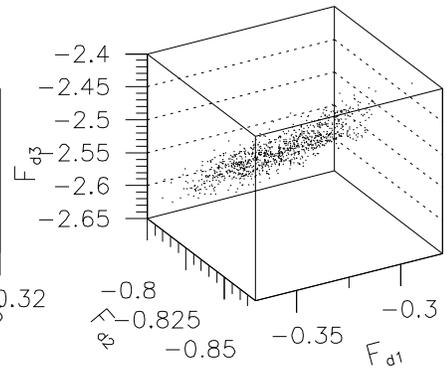
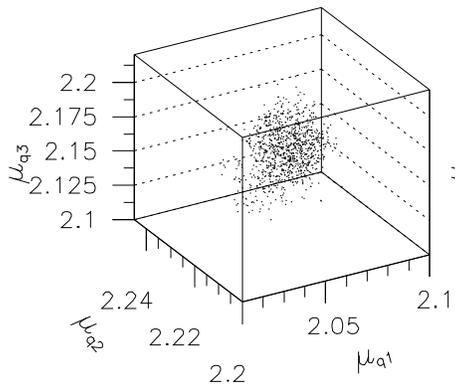
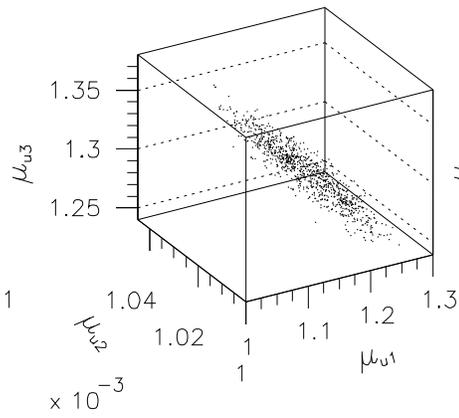
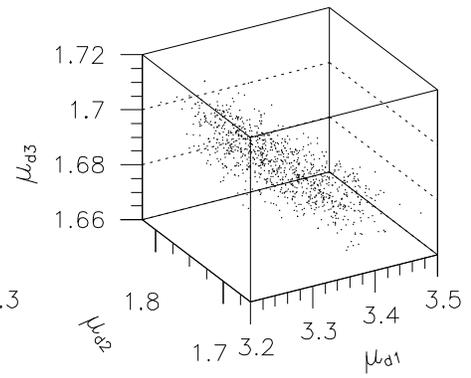
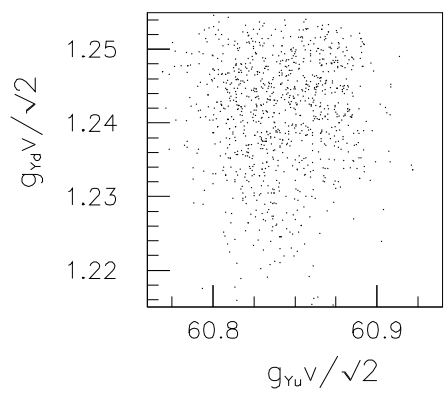
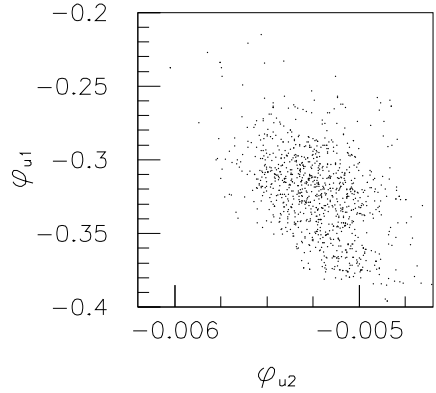
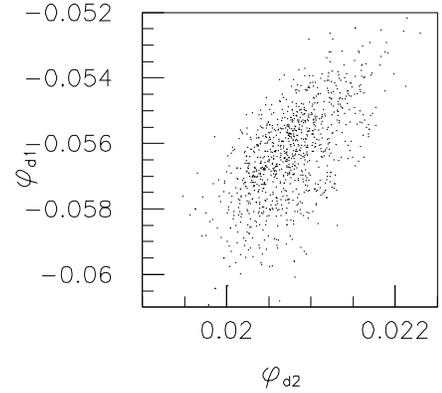
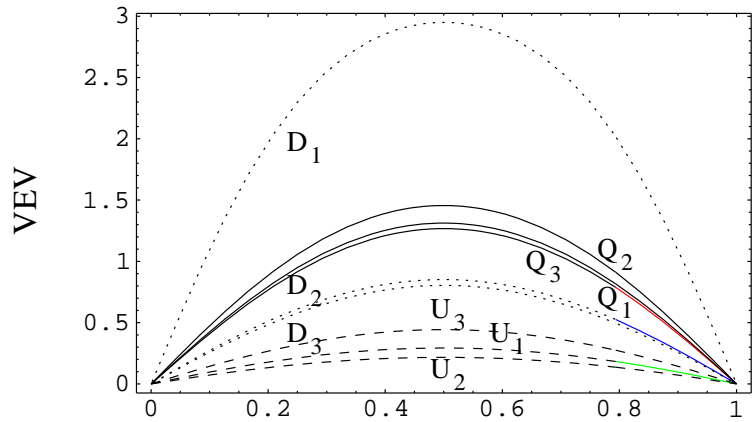
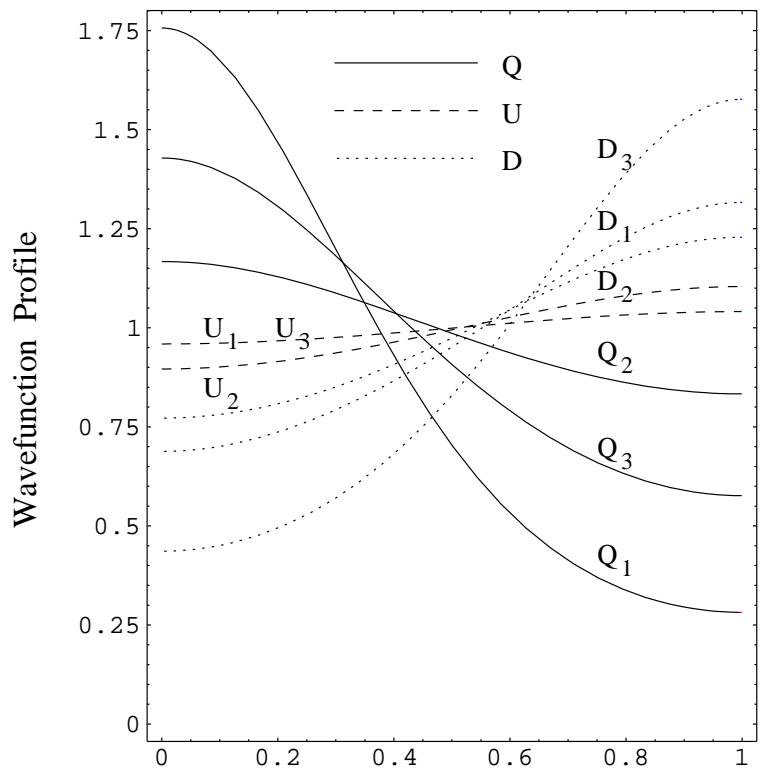
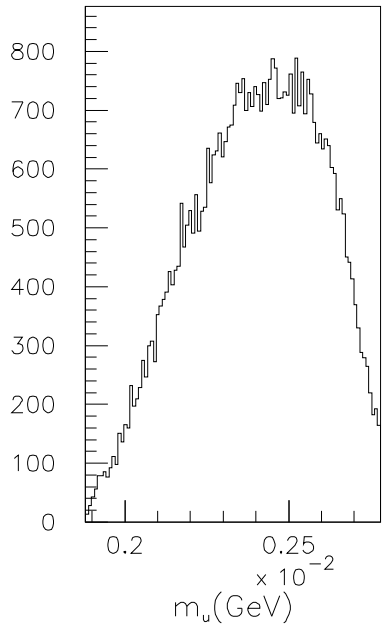
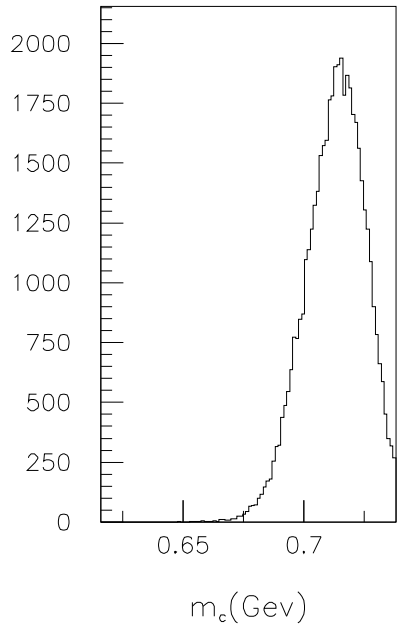
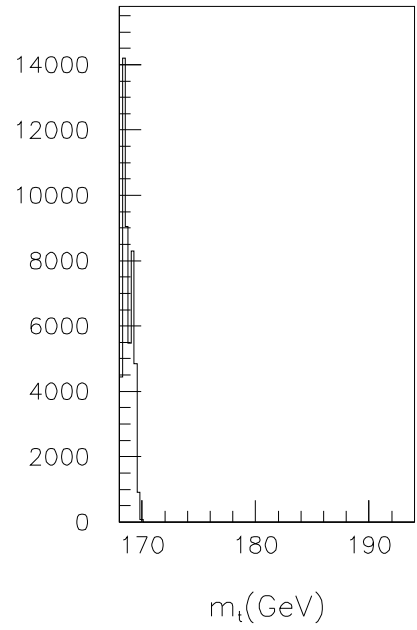
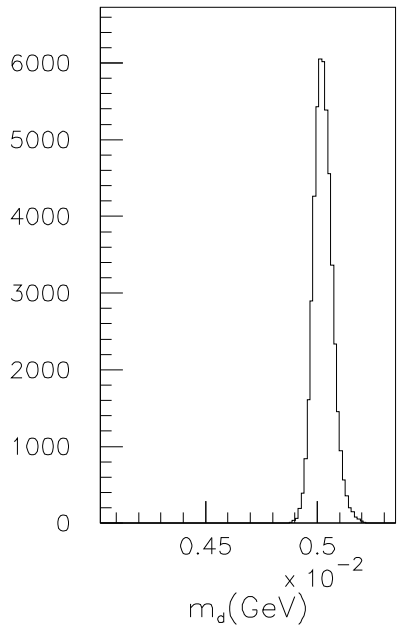
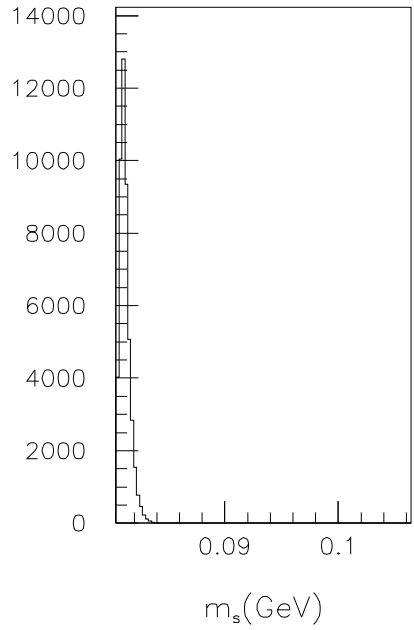
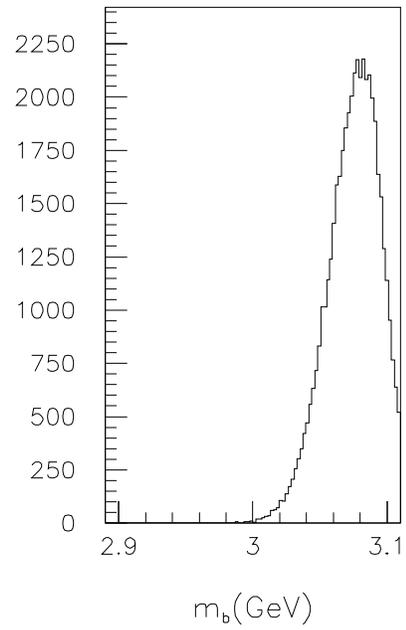
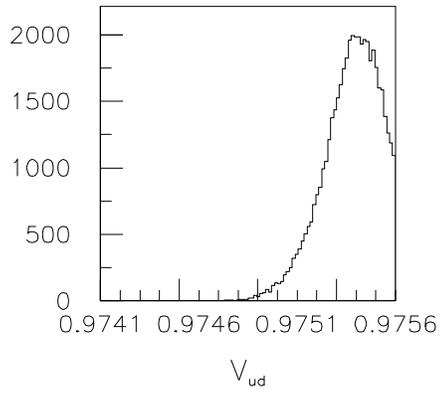
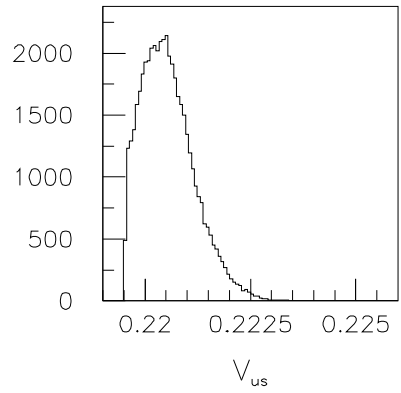
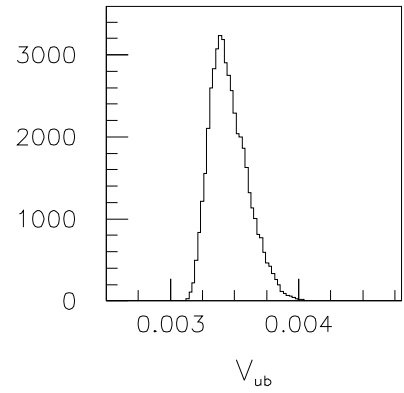
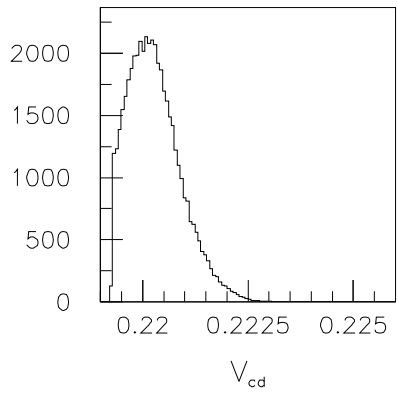
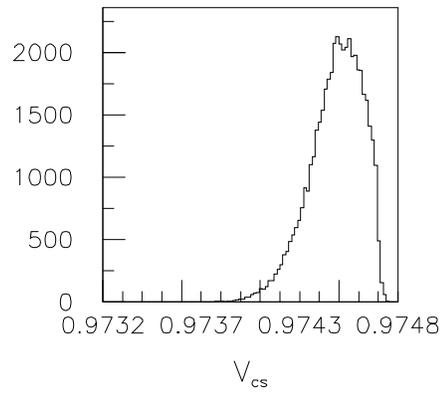
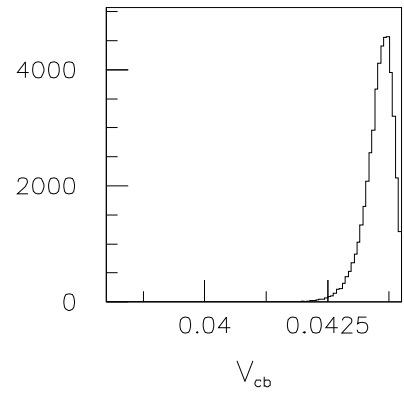
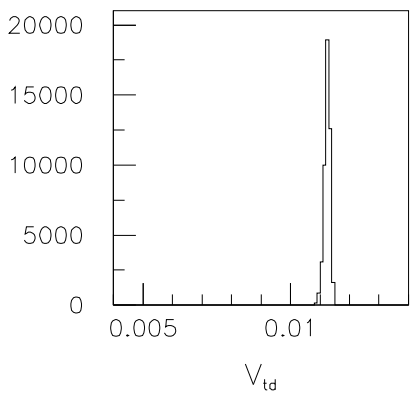
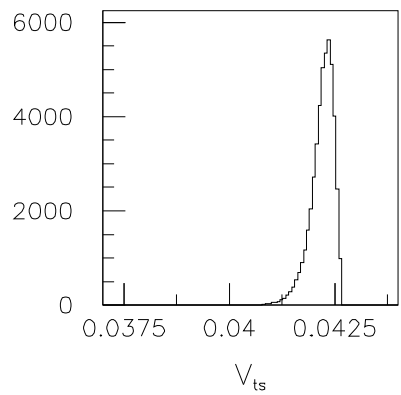
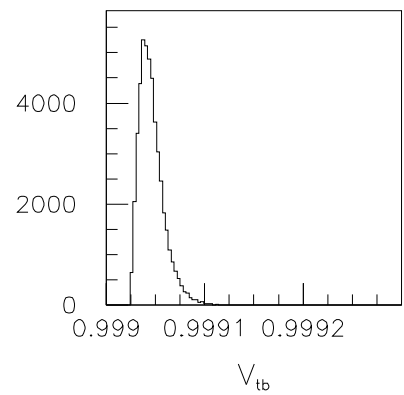
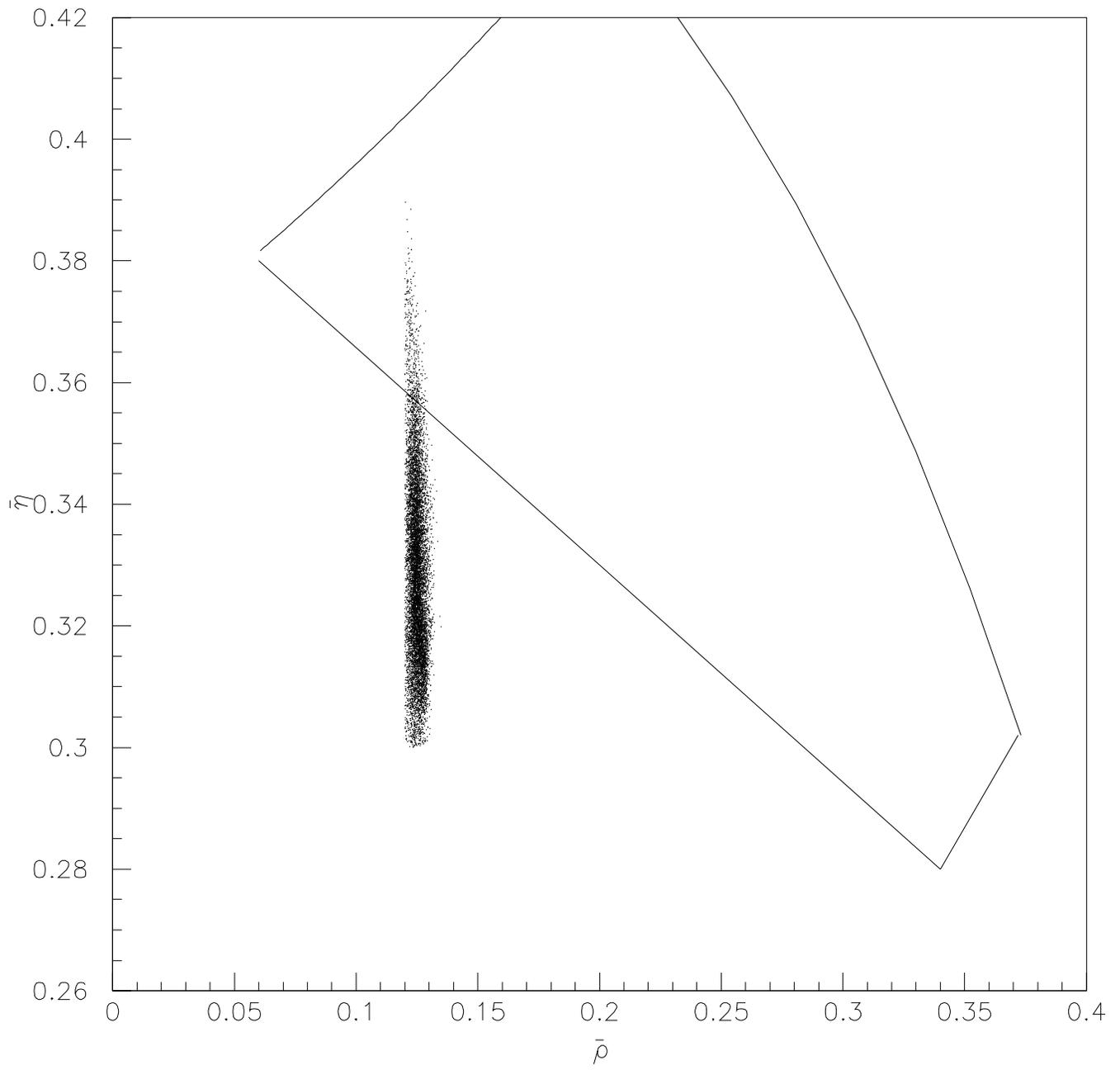
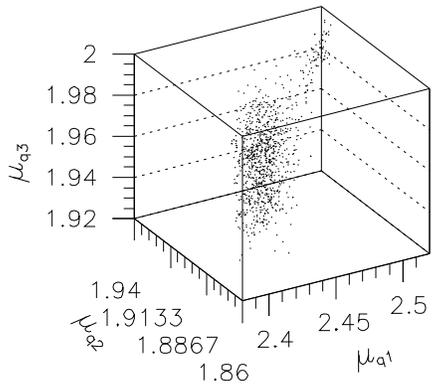
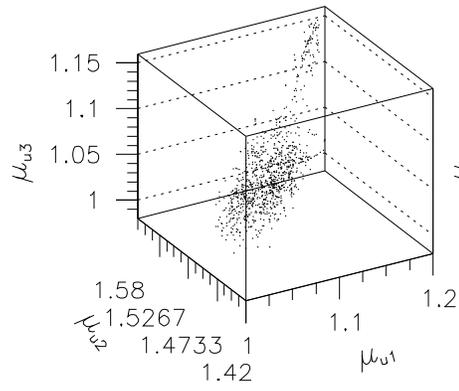
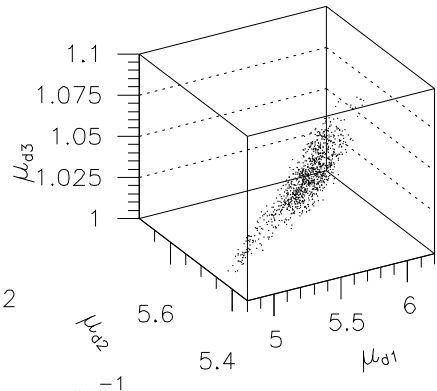
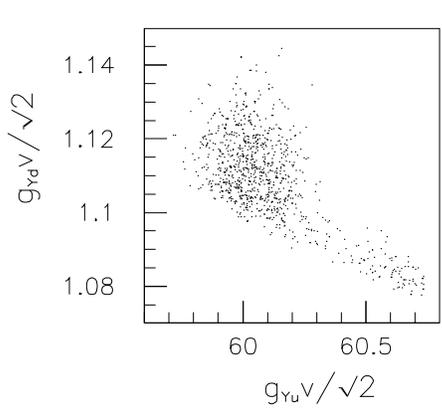
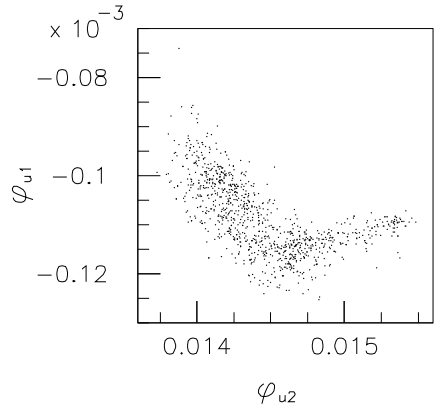
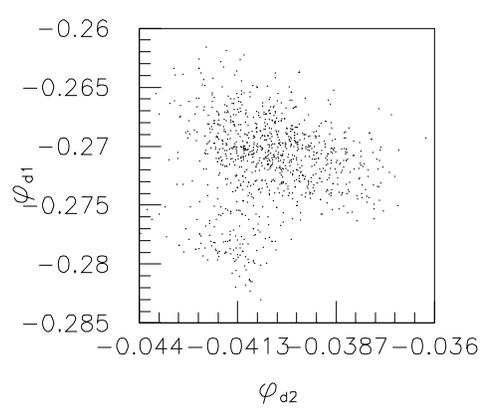
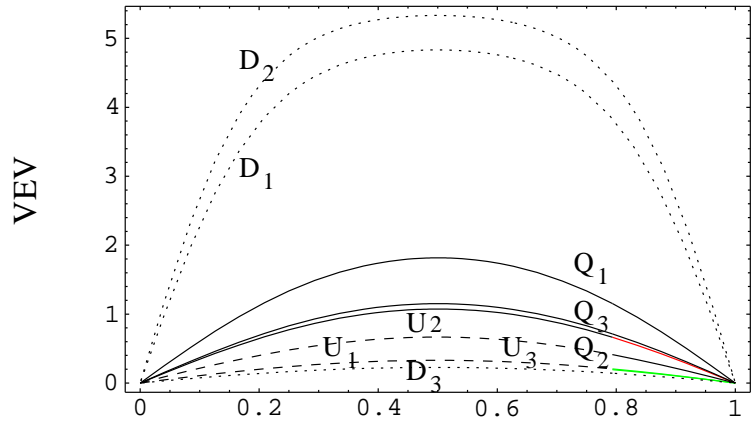
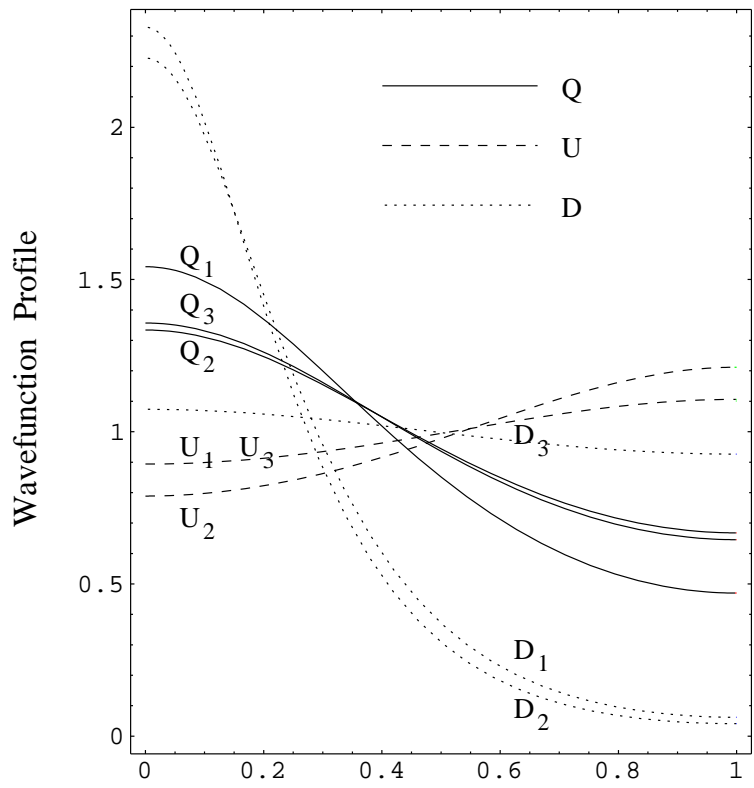
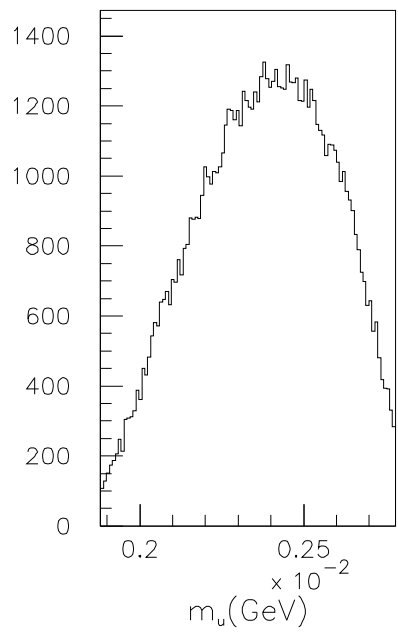
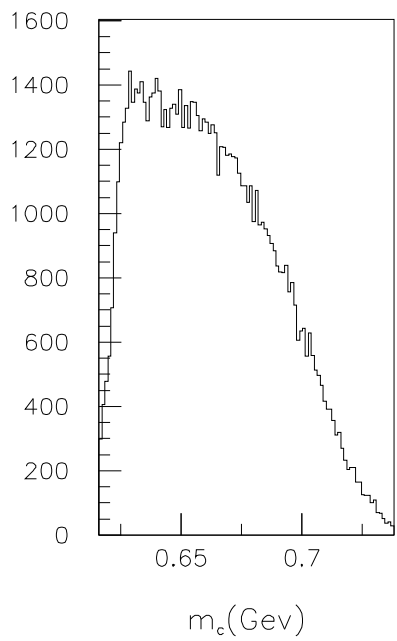
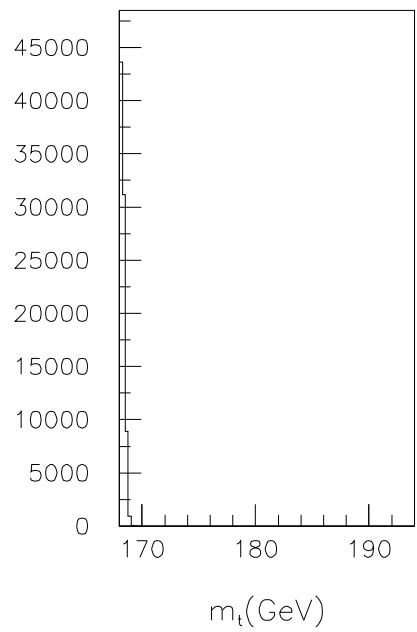
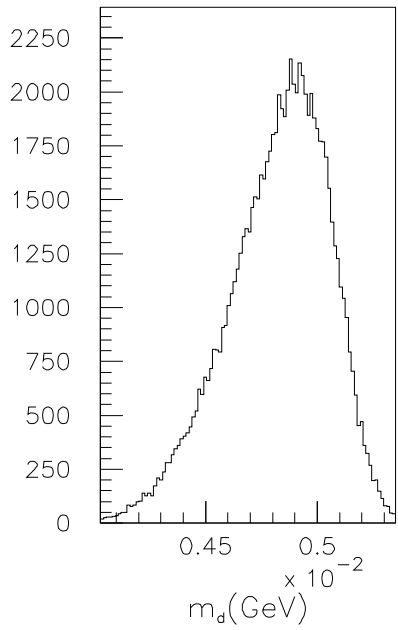
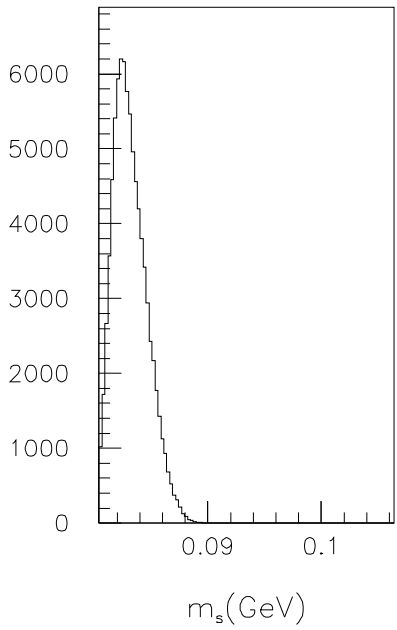
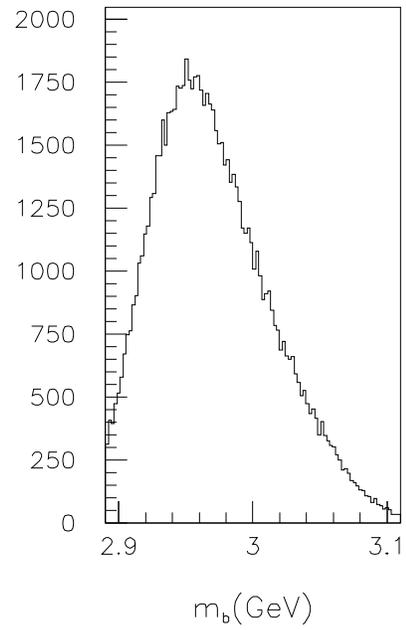
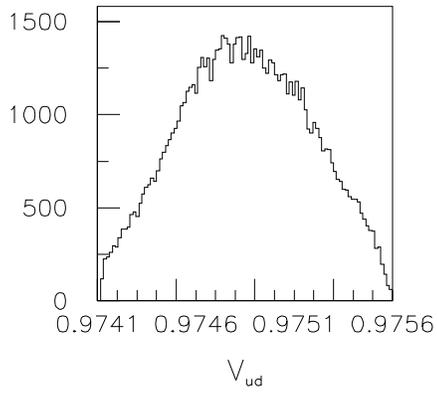
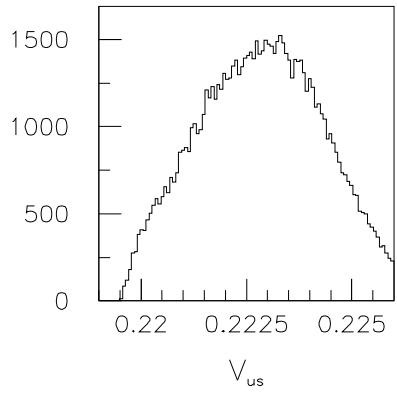
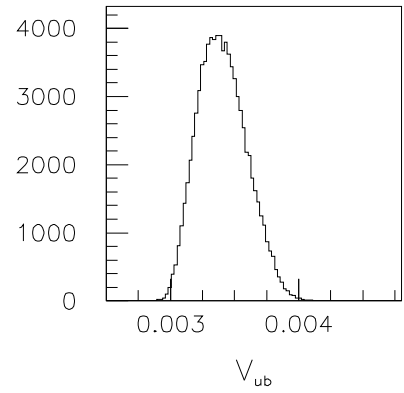
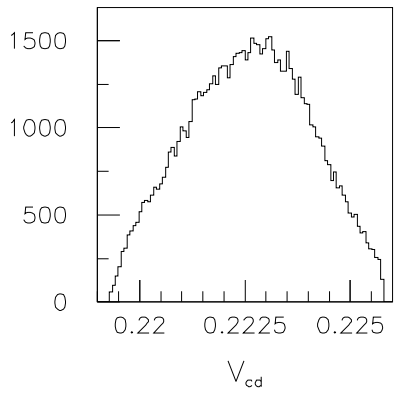
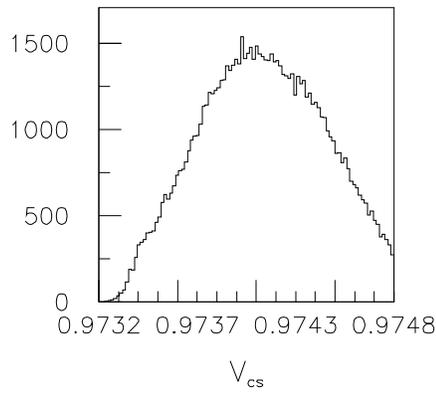
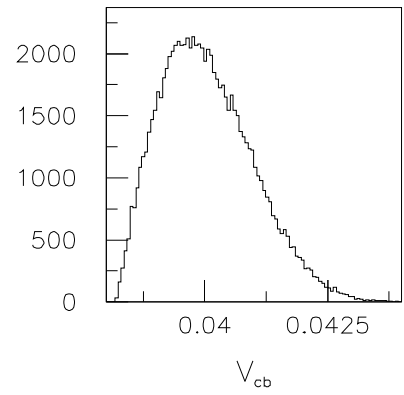
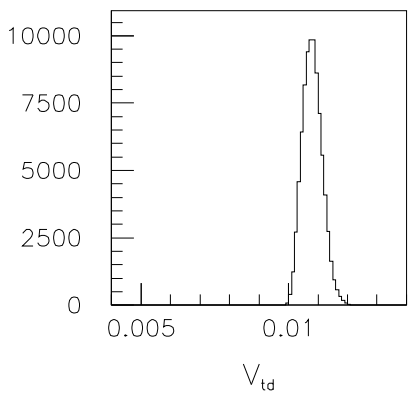
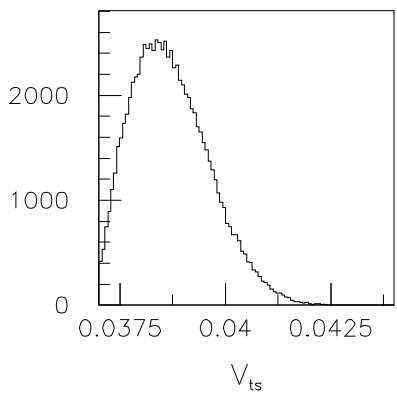
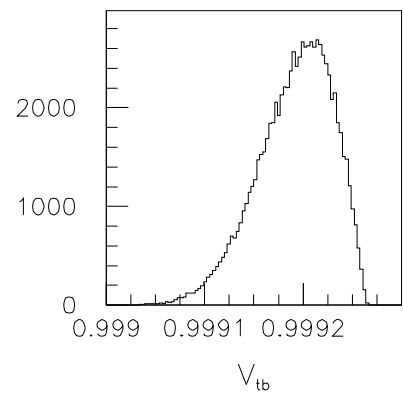
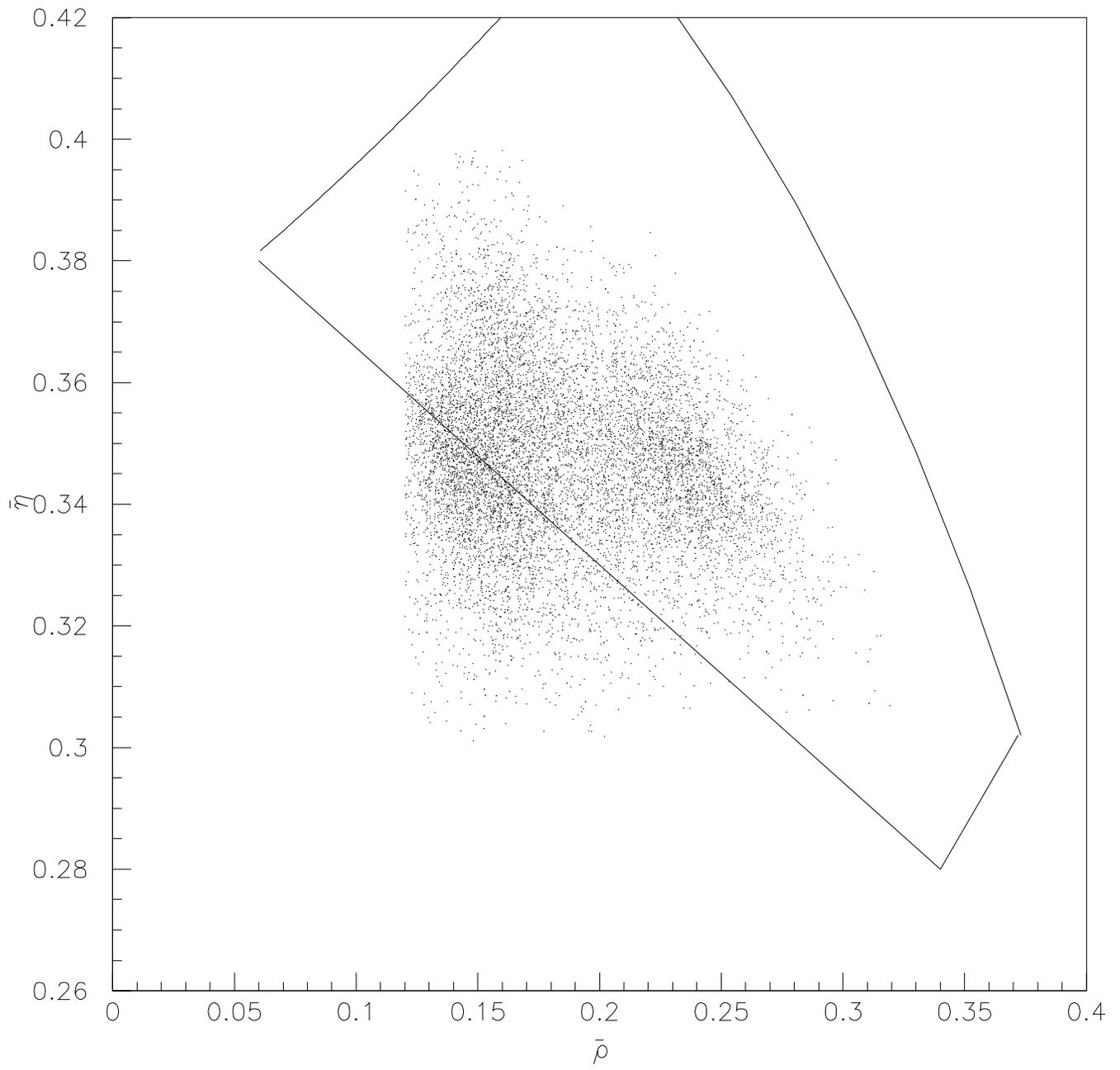
\end{document}